\documentclass[useAMS,usenatbib]{mnras}
\usepackage{times}
\usepackage{graphicx}
\usepackage{float}
\usepackage{epsfig}
\usepackage{courier}
\usepackage{chngcntr}
\usepackage{caption}
\usepackage{subcaption}
\usepackage{xspace}
\usepackage{epstopdf}
\usepackage{booktabs}
\usepackage{gensymb}
\usepackage{color,soul}
\usepackage{amssymb}
\usepackage{amsmath}
\usepackage[T1]{fontenc}
\usepackage{aecompl}
\usepackage[shortlabels]{enumitem}
\setlist[enumerate, 1]{1\textsuperscript{o}}
\usepackage{nicefrac}
\usepackage{rotating} 
\usepackage{tikz}
\usepackage[T1]{fontenc}
\usepackage{times}
\usepackage{graphicx}
\usepackage{textcomp}

\newcommand{\mj}{M$_{\mathrm{J}}$\xspace}
\newcommand{\msun}{$\mathrm{M_{\odot}}$\xspace}

\counterwithout{figure}{section}

\graphicspath{ {./figures/} {./figures/0.6msun/}} 

\title[The morphology of Elias 2-27]{Is the spiral morphology of the Elias 2-27 circumstellar disc due to gravitational instability?}
\author[Hall et al.]{Cassandra Hall$^{1,2}$\thanks{Email: cassandra.hall@le.ac.uk}, Ken Rice$^{2,3}$, Giovanni Dipierro$^{1}$, Duncan Forgan$^{4,5}$,Tim Harries$^{6}$ \newauthor and Richard Alexander$^{1}$\\
$^{1}$Department of Physics \& Astronomy, University of Leicester, Leicester, LE1 7RH, UK \\
$^{2}$SUPA\thanks{Scottish Universities Physics Alliance}, Institute for Astronomy, University of Edinburgh, Blackford Hill, Edinburgh, EH9 3HJ, UK\\
  $^{3}$Centre for Exoplanet Science, University of Edinburgh, Edinburgh, UK\\
$^{4}$SUPA\footnotemark[2], School of Physics \& Astronomy, University of St Andrews, North Haugh, St Andrews, KY16 9SS, UK\\
$^{5}$St Andrews Centre for Exoplanet Science, University of St Andrews, UK\\
$^{6}$University of Exeter, Stocker Road, Exeter EX4 4QL, UK\\
}
\begin{document}

\date{\today}

\pagerange{\pageref{firstpage}--\pageref{lastpage}} \pubyear{2016}

\maketitle

\label{firstpage}

\begin{abstract}
A recent ALMA observation of the Elias 2-27 system revealed a two-armed structure extending out to $\sim$300 au in radius. The protostellar disc surrounding the central star is unusually massive, raising the possibility that the system is gravitationally unstable. Recent work has shown that the observed morphology of the system \emph{can} be explained by disc self-gravity, so we examine the physical properties of the disc necessary to detect self-gravitating spiral waves. Using three-dimensional Smoothed Particle Hydrodynamics, coupled with radiative transfer and synthetic ALMA imaging, we find that observable spiral structure can only be explained by self-gravity if the disc has a low opacity (and therefore efficient cooling), and is minimally supported by external irradiation. This corresponds to a very narrow region of parameter space, suggesting that, although it is possible for the spiral structure to be due to disc self-gravity, other explanations, such as an external perturbation, may be preferred.
\end{abstract}

\section{Introduction}
\label{sec:intro}
It has long been known that spiral structure exists in galaxies, with many examples of striking spiral arms, such as those found in the Pinwheel galaxy (M101) and the Whirlpool galaxy (M51). 

Despite many examples of spiral structure in numerically simulated protostellar discs present in the literature (see, e.g. \citealt{boss1989,boss1998,gammie2001,riceetal2003}), these structures, unlike their galactic cousins, have only recently been clearly observed in nature (see, e.g.,  \citealt{garufietal2013,perezetal2014,benistyetal2015,stolkeretal2016}), due to significant advances in imaging capability.

One such advance has been SPHERE \citep{sphereinstrument}, the Spectro-Polarimetric High-contrast Exoplanet REsearch instrument mounted on UT3 at the VLT. Using extreme adaptive optics coupled with a coronograph and polarimetric differential imaging, we are able to obtain images of unprecedented resolution and sensitivity of the scattering surface of protostellar discs \citep{benistyetal2015,stolkeretal2016}. 

Since the Atacama Large Millimeter/submillimeter array (ALMA) began operations, the astronomical community has, for the first time, been able to resolve midplane structure in protostellar discs. With this capability has come an array of unexpected results - these discs are often far from smooth, and a plethora of substructures have been revealed. These include features such as the multiple rings clearly observed in the HL Tau \citep{hltau} and TW Hydrae \citep{twhydrae} systems. It is possible that these rings are caused by planets carving gaps in the dust of the disc \citep{dipierrohltau,jinetal2016}, but these features also have plausible alternative explanations, such as aggregate sintering \citep{okuzumietal2016}, particle trapping at the edge of the disc dead-zone \citep{rugeetal2016} and particle concentration at planet-induced gap edges \citep{zhuetal2014}, to name a few.

In the earliest stages of its lifetime, it is likely that an accretion disc surrounding a central protostar will be massive enough to be self-gravitating (e.g., \citealt{linpringle1990}). Consequently, the gravitational instability may play an important role in the early evolution of such a system. For a protostar to accrete from its nascent disc, angular momentum must be redistributed outwards, so that mass may move in while conserving angular momentum. If the disc is self-gravitating, the gravitational instability (GI) is largely responsible for this angular momentum transport \citep{cossinslodatoclarke2009, forganetal2011}.

However, a disc in Keplerian rotation is only gravitationally unstable to axisymmetric perturbations if the Toomre parameter, $Q$, is \citep{toomre1964}
\begin{equation}
Q = \frac{c_\mathrm{s}\kappa}{\pi\mathrm{G}\Sigma} \lesssim 1,
\end{equation}
where $c_\mathrm{s}$ is the sound speed, $\kappa$ is the epicyclic frequency (which in a Keplerian disc is simply $\Omega$, the orbital frequency), G is the gravitational constant and $\Sigma$ is the surface density of the disc. For non-axisymmetric perturbations, this value has been determined empirically as $Q\approx1.5-1.7$ \citep{durisenetal2007}. The Toomre criterion is a local quantity, real protostellar discs are likely to be self-gravitating in some regions (i.e., at large radii) and not in others. However, this can be translated to a global condition for instability, that the disc-to-star mass ratios, $q$, satisfies 
\begin{equation}
q=\frac{M_\mathrm{d}}{M_*}\sim \frac{H}{R} \gtrsim 0.1
\end{equation}
for a disc in Keplerian rotation in vertical hydrostatic equilibrium.
Surveys of protostellar discs, such as those conducted by \citet{andrewsetal2013} of the Taurus star-forming region, likely show us that by the ~Class II stage, these objects are not massive enough for disc self-gravity to be important. However, there are certainly examples where it may indeed be important \citep{andrewsetal2009}.

One such example is the Elias 2-27 system, classified as a Class II young stellar object. It has an unusually large disc-to-star mass ratio (measured assuming a gas-to-dust ratio of 100:1), $q\sim 0.24$ \citep{andrewsetal2009}. Inherent uncertainties in the physical properties of the dust grains from which mass is inferred, such as opacity, size distribution, and dust mass fraction, mean that there are significant uncertainties attached to estimates obtained this way. Furthermore, self-gravitating discs are expected to be optically thick at $\sim$sub-mm wavelengths, so large amounts of hidden mass may be screened. However, this is dependent on the nature of the grain distribution. For example, if grain growth has occurred, such that more of the dust mass is concentrated in larger grains within the grain distribution, then we may, in fact, expect the disc to remain optically thick out to $\sim 3$mm wavelengths \citep{forganrice2013}, which will further conceal hidden mass. Even discounting this, it does appear that Elias 2-27 is sufficiently massive to be self-gravitating.

A self-gravitating disc in Keplerian rotation will develop spiral structure, since the disc becomes unstable to non-axisymmetric modes that grow non-linearly at $Q\lesssim 1.5$ \citep{papaloizousavonije1991,durisenetal2007}. Once these modes have developed, the disc will settle into a quasi-steady, self-regulating state where its $Q$ value remains roughly constant. In a sense, self-gravitating discs afre more readily unstable to these non-axisymmetric modes, since the development of axisymmetric modes requires $Q\sim 1$. While there are multiple examples of spiral features observed in scattered light (see, e.g. \citealt{benistyetal2015,mutoetal2012}), at the wavelengths ($\sim$ micron) of these observations the majority of the disc is optically thick. As such, only the surface of the disc is observed. To reveal midplane structure requires observations at optically thin wavelengths.

The Elias 2-27 system was the first system probed down to the disc midplane that revealed spiral structure \citep{perezetal2016}. Observed using ALMA at 1.3 mm, the system was spatially resolved, revealing two spiral arms extending in radius out to 300 au from the central star. With such a large disc mass, it is possible that this is the first observed example of gravitational instability in a protostellar disc. However, it is not the only system where this appears to be the case. The Class 0 triple protostar system L1448-IRS3B is thought to have formed through fragmentation due to gravitational instability \citep{tobin2016}.

If a protostellar system is self-gravitating, and exists in a quasi-steady state of self-regulation,  then the number of spiral arms, $m$, is related to the disc-to-star mass ratio, $q$, by $m\sim1/q$ \citep{cossinslodatoclarke2009,donghallrice2015}. If the measured disc mass is correct for the Elias 2-27 system, we may expect 4 spiral arms, rather than 2. Conversely, if the system is self-gravitating, then to obtain a 2 armed spiral would require $q\sim 0.5$. External perturbers, such as stellar flybys, can also produce this classical two armed spiral structure \citep{quillenetal2005}. It is important to note that it may be possible for a higher number of spiral arms to be present than are actually observed (see, e.g., \citealt{dipierro2014}). 

Protostellar discs neither form, nor evolve, in isolation. Particularly at very early times, they are heavily embedded in their nascent cloud of formation, and undergo considerable infall from this cloud. Indeed, observed accretion rates onto protostellar discs are typically an order of magnitude larger than observed accretion rates onto the stars themselves (see, e.g. \citealt{kenyonetal1990,calvethartmannstrom2000}). This has a significant effect on the dynamical evolution of the system, driving power into the lower order, global spiral modes (i.e. $m\sim 2$), resulting in increased efficiency of angular momentum transport \citep{harsonoetal2011}. The issue is further complicated by the resolution of the observation; systems which are dominant in the high $m$-modes may appear to have fewer spiral arms than are actually present, due to size scales (i.e., spiral arm widths) that might be smaller than the beam size of the observation \citep{dipierro2014}.

This results in a narrow region of parameter space where spirals due to disc instability may be detected by an instrument such as ALMA, an idea examined in the semi-analytic parameter space investigation of \cite{halletal2016}. This semi-analytic sweep of parameter space focused only on the \emph{local} angular momentum transport case, which in 3D global hydrodynamics simulations using $\beta$ cooling \citep{lodatorice2004} has been shown to be valid for disc-to-star mass ratios of $q\lesssim 0.25$. \citet{halletal2016} found that in all cases of spirals generated due to gravitational instability under local angular momentum transport, the relatively small spiral amplitudes (when compared to global spirals) were difficult to detect using ALMA. 

A logical deduction, therefore, is that any spirals generated by the gravitational instability, that are detectable (and resolved fully, i.e., 2 arms present, 2 arms detected) by ALMA, may be spirals that are transporting angular momentum \emph{globally}, and, as such, will not be well-described by a semi-analytic formalism. For local angular momentum transport, the spirals must exist in what is known as the ``tight-winding approximation'', \citep{binneytremainedynamics} which removes long-range coupling forces by assuming the waves are tightly wound. The pitch angle, $i$, is given by
\begin{equation}
\tan i = \bigg| \frac{m}{kR}\bigg|,
\end{equation}
where $k$ is the radial wavenumber, $R$ is the radial distance from the central star, and $m$ is the number of spiral arms. For the tight-winding approximation to be valid, we require
\begin{equation}
\label{eq:wkbvalid}
\bigg|\frac{m}{kR}\bigg| \ll 1.
\end{equation}
Under these conditions, radial spacing between each spiral arm is small, since the radial distance between waves, $\Delta R$, is given by
\begin{equation}
\Delta R = \frac{2\pi}{k},
\end{equation}
and the condition in \ref{eq:wkbvalid} implies that $k$ is large. This small radial spacing between spiral arms undergoing local angular momentum transport makes them more difficult to resolve than spiral arms that transport angular momentum globally. Recent numerical work by \citet{meruetal2017} and \citet{tomidaetal2017} has shown that the Elias 2-27 morphology \emph{can} be caused by the gravitational instability. In this work, we take a broader approach, performing hydrodynamical simulations of self-gravitating protostellar discs to determine the region of parameter space in which we expect self-gravitating spirals to be observable.  

\section{Method}
\label{sec:method}
We run a suite of 17 Smoothed Particle Hydrodynamics (SPH) simulations of a protostellar disc, varying only the disc mass, metallicity, and minimum irradiative background temperature of each disc. Our intention is to scan the parameter space that the Elias 2-27 system may exist in. We vary the disc-to-star mass ratio between $q=0.25$ and $q=0.5$, vary the minimum irradiative background temperature between $T_{\mathrm{min}}=5-15$ K, and assume either 1.0 $\times$ solar metallicity or $0.25\times$ solar metallicity. Each system is evolved until it \textit{either}:
\begin{enumerate}[(1),leftmargin=1\parindent]
\item reaches a marginally unstable (since it is partially supported by external irradiation), self-gravitating state in which spiral arms are present, \textit{or}
\item fragments, \textit{or}
\item becomes clear that no non-axisymmetric structure is going to develop (e.g. after $\sim 20$ orbital rotation periods, which corresponds to $T\sim 50,000$ years).
\end{enumerate}


In addition to this, we present results where the spiral structure has been amplified by the method outlined in section \ref{sec:amplify}. This enhancement may be considered as a proxy for solid particle trapping in the spiral arms. Gas in a protostellar disc usually orbits at slightly sub-Keplerian velocity since it is partially supported by an outward pressure gradient. However, solid particles do not feel pressure, so continue to orbit at Keplerian velocity, resulting in a headwind. This headwind produces a drag force on the solid particles, causing them to lose angular momentum and drift radially inwards.

However, self-consistently simulating grains of these sizes in a hydrodynamics simulation of a self-gravitating disc has proved to be problematic. The conventional approach has been to model two fluids, coupled via a drag term (c.f., \citealt{riceetal2004}). However, this method is not best suited for grain sizes that are of the most interest to us, since very small timesteps are required to capture the motion of small grains, resulting in prohibitively long integration times. 

%
 
%

For this reason, rather than expend significant computational resources simulating these grains for a suite of hydrodynamical models, we use the spiral amplification method outlined  in section \ref{sec:amplify} as an approximation for dust-trapping. We present these results together with non-amplified discs in section \ref{sec:results}.
 
In all cases where spiral structure develops, we use the \texttt{TORUS} radiation transport code \citep{harriesetal2004,kurosawaetal2004,haworthetal2015} to perform a radiative transfer calculation. We feed this into the ALMA simulator in \texttt{CASA} (ver 4.7.2; \citealt{CASA}) to produce a synthetic ALMA observation, with the same parameters (i.e. wavelength, integration time, antenna configuration, precipitable water vapour) as the original observation in \citet{perezetal2016}. Each synthetic observation is then processed using the same unsharp masking filter image processing technique as described in \citet{perezetal2016}, which we explain here in section \ref{sec:unsharp}.

\subsection{Smoothed Particle Hydrodynamics}
\label{subsec:SPH}

Smoothed Particle Hydrodynamics (SPH) is a Lagrangian method for simulating hydrodynamical systems, where the fluid is discretised into pseudo-particles \citep{lucy1977,gingoldmonaghan1977}. There are many reviews available that describe SPH (see, e.g. \citealt{rosswog2009, monaghan1992}), so we simply give an overview of the techniques used in this work.

Each disc is modelled by $N$ pseudo particles, and each pseudo-particle has a mass, position, velocity and internal energy. The density is obtained by interpolation, giving all the properties of the fluid at a given location.

Pressure is obtained by using the internal energy of the particle coupled with an equation of state. Gravitational forces are calculated using \texttt{TREE} algorithms \citep{barneshut1989}, and the discretised equations of energy and momentum are then solved. Particle velocities are updated using these pressure and gravitational forces, and particle positions updated using these velocities. Internal energy changes are calculated from radiative cooling and heat conduction, viscous dissipation and $P\,\,\mathrm{d}V$ work.

%

Fully polychromatic radiative hydrodynamics, while possible, \citep{acremanharriesrundle2010}, is prohibitively computationally expensive, especially for self-gravitating systems. To overcome this, we use the hybrid radiative transfer method of \citet{forgan2009}. Here, the polytropic cooling approximation of \citet{stamatellos2007}, which takes account of local optical depth, is combined with the flux-limited diffusion method of \citet{mayeretal2007}, which models energy exchange between particles. In this manner, we have, perhaps, a ``closer" approximation to the complexity of the real radiative processes involved than with a parameterised cooling prescription (such as the so-called $\beta$-cooling prescription, see \citealt{gammie2001,rice2003}) alone.

\subsection{SPH simulation setup}
\label{subsec:sphsetup}
We ran a total of 17 SPH simulations, each with a resolution of 2 million particles. We report a ratio of smoothing length to scale height of $h/H <0.25$ beyond $\sim 50$ au, satisfying the resolution requirements for SPH simulations of circumstellar discs laid out in \citet{nelson2006} (i.e., $\sim 4$ smoothing lengths per scale height ). Parameters either match or nearly match
those of the of the disc in the Elias 2-27 system. The central star mass in
Elias 2-27 is $M_*=0.6$ \msun \citep{andrewsetal2009,nattaetal2006}, and
spiral structure is observed in the continuum out to $r=300$ au
\citep{perezetal2016}, for this reason we run the SPH simulations with a disc
outer radius of $R_{\mathrm{out}}=300$ au. However, this could be a lower
limit to the radial extent of the disc. Since continuum emission traces dust,
and dust undergoes inward radial migration, it is feasible that the radial extent of the gas disc could be larger than that of the dust disc (e.g., \citealt{perezetal2012}).

The issue of disc mass is similarly complicated. The observed disc mass (inferred from dust mass) of Elias 2-27 is $\sim 0.04-0.14$ \msun \citep{andrewsetal2009,isellaetal2009,riccietal2010}, which gives a disc-to-star mass ratio, $q$, of somewhere in the range $q\sim 0.06-0.23$. Typically, disc self-gravity only becomes important for $q\gtrsim 0.1$, and the number of spiral arms, $m$, we expect to be present in a self-gravitating disc is related to $q$ through $m\sim 1/q$ \citep{cossinslodatoclarke2009,donghallrice2015}. If the spiral structure in Elias 2-27 is due to disc self-gravity, and the estimated disc mass is correct, then we may expect this system to have 4 spiral arms.


 However, the observation of \citet{perezetal2016} shows a \textit{two}-armed spiral, which would typically require a system with $q\sim 0.5$ \citep{lodatorice2005,cossinslodatoclarke2009,donghallrice2015}. However, the empirically determined $m\sim 1/q$ relationship is not exact. For example, Fourier decomposition of the mode spectrum can show approximately equal power in neighbouring modes (see Figure 2 of \citealt{lodatorice2004}), so may be dominant in higher $m$-modes than $m=2$. If this is the case for Elias 2-27, it has been shown that discs may be observed to have fewer arms than are present in reality \citep{dipierro2014}.
 
 Such a massive, extended disc (i.e., $q\sim 0.5$, $R_{\mathrm{out}}\sim 300$ au) is likely to be susceptible to fragmentation, unless it is partially supported by some external irradiation \citep{levin2007,clarke2009,rafikov2009,halletal2016}. In the case of Elias 2-27, this is physically motivated by the system being surrounded by its parent's molecular cloud \citep{nattaetal2006}. This cloud creates a temperature bath, setting the minimum disc temperature to order 10 K \citep{casellimyers1995}.
%

\begin{table*}
\centering
\begin{tabular}{|c|c|c|l|c|c|c|}
\hline
\textbf{Mass ratio}   & \textbf{Metallicity (solar) }    & \textbf{${\mathbf{T}}_{\mathrm{\min}}$} & \textbf{Outcome}  &  \textbf{Figure} & \textbf{Detect spirals?} & \textbf{Detect spirals if amplified?}\\
\hline
\hline
0.25   & 0.25    &  5            &  Smooth                                               &  - & - & -\\  
0.25   & 1.0                           &  5           &  Smooth                                               &  - & - & -\\  
0.325 &  0.25   &  5            &  \textbf{Spirals}                                  &    \ref{fig:lowmassdiscs}  & Yes & Yes, Fig \ref{fig:lowmassamp}   \\ 
0.325 &  0.25  &  10          &  Smooth                                              &  - & - & -  \\ 
0.325 &  0.1     &   5           &  Spirals in inner disc,                        &   - & - & - \\
           &                                       &                   & smooth outer disc                              &         \\
0.325 & 1.0                        &   5            &  \textbf{Spirals}                                  &    \ref{fig:lowmassdiscs}     & No & Yes, Fig \ref{fig:lowmassamp}\\  
0.325 &  1.0                        &   10         &  \textbf{Spirals}                                  &  \ref{fig:lowmassdiscs}       & No & No\\  
0.325 & 1.0                          &   15          &  Smooth                                              &   -  & - & -\\  
 0.4    &  0.25  &   5            &  \textit{Fragmented}                           &   -  & - & -\\  
 0.4    & 0.25    &   10          &  \textbf{Spirals}                                 &     \ref{fig:highmassdiscs}   & No & Yes, Fig \ref{fig:highmassamp}   \\  
 0.4    &  0.25  &   15          &  Smooth                                              &   -   & - & - \\  
 0.4    &  1.0                         &   5            &  \textit{Fragmented}                          &    -    & - & -\\  
 0.4    & 1.0                             &   7.5        &  \textit{Fragmented}                          &     - & - & -  \\  
 0.4    &  1.0                            &   10         &  \textbf{Spirals}                                 &      \ref{fig:highmassdiscs}   & No & Yes, \ref{fig:highmassamp}    \\  
0.4     &  1.0                            &   15         &  Smooth                                             &    -  & - & - \\  
 0.5    & 1.0                             &   10         &  \textbf{Spirals}                                 &     \ref{fig:highmassdiscs}  & No & Yes \ref{fig:highmassamp}    \\
 0.5    & 1.0                             &   15         &  Smooth                                             &    -  & -& -   \\
 \hline
\end{tabular}%
\caption{Table of all simulations conducted in this work. Columns from left to right are: disc-to-star mass ratio, metallicity (assuming 1:1 relationship with opacity), minimum temperature floor, the outcome of the simulation (whether spirals due to gravitational instability are present or not), the corresponding Figure number for each simulation, whether spirals are visible in the image, and finally if they are visible in the image after amplification. No images were produced for discs that did not have observable spirals, or that fragmented.\label{tab:simulations}}

\end{table*}

With this in mind, we ran a suite of simulations with a minimum temperature to support this disc in a marginally unstable state. The details of each simulation are shown in Table \ref{tab:simulations}. The central star, in all simulations, is $M_*=0.6$ \msun, inner disc radius is $R_{\mathrm{in}}=10$ au and outer disc radius is $R_{\mathrm{out}}=300$ au. The initial disc surface density profile takes the form $\Sigma\propto r^{-1}$, where $\Sigma$ is surface density and $r$ is disc radius, and the initial sound speed profile is $c_\mathrm{s}\propto r^{-0.25}$. The minimum temperature is set between 5 K and 15 K. The disc-to-star mass ratio is varied from $q=0.25$ to $q=0.5$, giving a total disc mass of between 0.15 \msun and 0.3 \msun. Evidence from observation of $\sim$3 Myr old systems \citep{andrewsetal2013} suggests that the disc masses we have simulated are approximately an order of magnitude more massive than disc masses that are observed. However, inferred disc masses from observations come with considerable uncertainties, primarily from the inherent uncertainty in the dust-to-gas mass ratio (see, e.g. \citealt{andrewswilliams2005},\citealt{williamsbest2014}). If the disc masses are, therefore, underestimated by a factor a few, it is possible that many of these discs will be sufficiently massive so as to be self-gravitating.

The metallicity is either $0.25\times$ solar, or solar; motivated by the derived Toomre parameter only being low enough for the disc to be gravitationally unstable if the 1.3 mm opacity is $0.25\times $ solar (see figure S2 in the supplementary material of \citealt{perezetal2016}). This can be equivalently stated as $0.25\times$ solar opacity is the highest possible opacity that still results in a Toomre parameter of $Q<2$.

Since opacity is a local quantity, for a given composition the relationship between metallicity and opacity is precisely determined \citep{semenovetal2003}. However, this precise determination does rely upon knowing the exact composition of the material that is providing the opacity. Physically, in observed protostellar discs, some standard assumption about composition is made so that opacities can be estimated.  We note that modifying the opacity in this way is, essentially, equivalent to modifying the fractional abundance of dust by mass (see, e.g., equations 1-5 in \citealt{pollackmckay1985}). 

 
 Numerical work (e.g., \citealt{boss2002,johnsongammie2003}) has shown that protostellar disc instability is relatively insensitive to changes in opacity, since the timescale for temperature equilibriation by radiative diffusion is \citep{laughlin1994}
\begin{equation}
\tau_{\mathrm{rad}} = \frac{\kappa\rho H^2}{c},
\end{equation}
where $\kappa$ is opacity, $\rho$ is density, $c$ is the speed of light and $H$ is the scale height of the disc. Typical values give $\tau_{\mathrm{rad}}=100\tau_{\mathrm{dyn}}$, where $\tau_{\mathrm{dyn}}$ is the dynamical timescale (or orbital period) of the disc. Since collisional heating occurs on dynamical timescales, and the dynamical timescale is two orders of magnitude shorter than the radiative diffusion timescale, it is probably reasonable to assume that dynamical processes, rather than radiative ones, determine the long-term evolution of the system. We do not, therefore, consider a large number of radiative parameters. Instead, we simply consider two cases: solar opacity and $0.25\times$ solar opacity.

To understand how opacity affects the cooling of the disc, we first describe its implementation in our SPH code. The polytropic (\citealt{stamatellos2007}) cooling rate, per unit mass, of particle $i$ is given by
\begin{equation}
\label{eq:dudt}
\frac{\mathrm{d}u_i}{\mathrm{d}t} = \frac{4\sigma (T_0^4(\mathbf{r}_i) - T_i^4)}{\bar{\Sigma}^2 \bar{\kappa}_i(\rho_i, T_i) + \frac{1}{\kappa_i}(\rho_i, T_i)},
\end{equation}
where $\sigma$ is the Stefan-Boltzmann constant, $T_0(\mathbf{r}_i)$ is the temperature of the radiation field at the location ($\mathbf{r}_i$) of particle $i$, $T_i$ is the temperature of particle $i$, $\rho_i$ is the density and $\bar{\kappa}_i$ is the mass-weighted opacity of the particle, defined as
\begin{equation}
\bar{\kappa}=\frac{\bar{\tau}}{\bar{\Sigma}},
\end{equation}
where $\bar{\tau}$ is the mass-weighted average of the optical depth. We can see from equation \ref{eq:dudt} that there is a smooth transition between optically thin and optically thick regimes, with maximum cooling efficiency when the optical depth is order unity (i.e., the photosphere). 

We assume that the relationship between opacity and metallicity is 1:1. If we are in the regime where the disc is optically \textit{thin} for a given metallicity, then increasing the metallicity, i.e., increasing the amount of solids in the disc, will cause the disc to cool more efficiently. If, on the other hand, we are in a regime in which the disc is optically \textit{thick} at a given metallicity, reducing the metallicity will increase the cooling efficiency. 





\subsection{Spiral amplification}
\label{sec:amplify}
Solid particles embedded in a protostellar disc undergo radial drift in the direction of a positive pressure gradient \citep{weidenschilling1977}, and, in the case of a self-gravitating disc, will experience significant trapping in the spiral arms \citep{riceetal2004,gibbonsetal2012,gibbonsetal2014,dipierro2015,boothclarke2016} where local pressure maxima occur. Particles collect at local pressure maxima, 
which results in particularly large enhancements of local particulate density. This trapping is maximised for particles with a Stokes number of
\begin{equation}
St = \frac{\tau_\mathrm{s}}{\tau_\mathrm{ed}} = 1,
\end{equation}
where $\tau_\mathrm{s}$ is the particle stopping time (i.e., the ratio between the momentum of the particle and the drag force acting on the particle), and $\tau_\mathrm{ed}$ is the eddy turnover time, determined by the characteristic length scale, $l_\mathrm{c}$, and characteristic velocity, $v_\mathrm{c}$, of the eddies,
\begin{equation}
\tau_{ed}=\frac{l_{\mathrm{c}}}{v_{\mathrm{c}}},
\end{equation}
which, in a self-gravitating disc, is $\tau_{ed}=\Omega_{\mathrm{K}}$, since $l_{\mathrm{c}}=H$ and $v_{\mathrm{c}}=c_{\mathrm{s}}$.

This has previously been explored for $\sim$metre sized planetesimals in self-gravitating discs (see, e.g.,\citealt{riceetal2004,riceetal2006}). For grains of these larger sizes (i.e. $\gtrsim 10$ cm), the opacity contribution to emission at 1.2 mm is at least an order of magnitude lower than grains of $\sim$mm sizes \citep{draine2006}. The majority of the opacity for $\sim$mm emission comes from $\sim$mm sized grains, with typical values of $\kappa$=1.2, 1.1 and 1.05 cm$^2$ g$^{-1}$ for 1 mm, 3 mm, and 1 cm sized grains respectively \citep{draine2006}.

Since $\sim$mm emission traces dust, even a small amount of trapping can significantly broaden the parameter space where spirals due to GI are detectable (Rice et al., in prep.), since the movement acts to increase the brightness contrast ratio at $\sim$mm wavelengths between arm and inter-arm regions.

We do not use an SPH code that separately integrates dust motions, due to the computational cost of doing so for a self-gravitating disc. In this limit, dust timesteps become increasingly small, and we wished to examine a wide range of parameter space. Instead, we employ a spiral amplification technique that, when we consider well-mixed gas and dust, effectively acts as a proxy to the dust-trapping we have described above. 

We begin by gridding the SPH data into $500\times 500$ cells in 2D ($r,\phi$), and at each radius calculate the azimuthally averaged surface density. We then iterate again, and in each cell, calculate whether the surface density is above or below the azimuthal average for this radius. This depends on whether the spiral arm is present or not. The basic idea is that in the spiral arms,
\begin{equation}
\label{eq:dsigma}
\Sigma_{\mathrm{spiral}} > \Sigma_{\mathrm{avg}},
\end{equation} 
where $\Sigma_{\mathrm{spiral}}$ is the surface density of the spiral arm and $\Sigma_{\mathrm{avg}}$ is the azimuthal average of the surface density at that radius. So long as the condition is satisfied in \ref{eq:dsigma}, then the spiral is present, and the mass and density are multiplied by some factor, $A_{\mathrm{initial}}$. In this work, we use one value for this factor, and set $A_{\mathrm{initial}}=10$. If the condition in equation \ref{eq:dsigma} is not satisfied, then we divide by the factor $A_{\mathrm{initial}}$. The spiral is then amplified, essentially by moving mass into the arm region and out of the inter-arm region. The resultant masses and densities are then normalised, so that the total disc mass before and after the amplification is conserved. In practice, this normalisation process reduces the actual value of the final amplification factor, $A$, since multiplying and dividing mass and density by the same factor will not conserve mass naturally. In all cases, $A\approx 1.15$ (to two decimal places) after normalisation. This results in a total enhancement of 15\%, deliberately chosen to match enhanced dust-to-gas ratios seen in SPH simulations of self-gravitating discs post-processed with a 1D dust dynamics code  (e.g., lower right panel of Figure 4 of \citealt{dipierro2015}). We do not exceed this value in order to examine the minimum amount of enhancement required to detect spirals comfortably in at least some of the discs. 

\subsection{Radiation transport code: TORUS}
\label{sec:torus}
We use the \texttt{TORUS} radiation transport code \citep{harriesetal2004,kurosawaetal2004,haworthetal2015} to calculate the spectral flux density of the SPH discs described above. It determines radiative equilibrium in a dusty medium by using the photon-packet Monte-Carlo method first described in \cite{lucy1999}. 

To perform this calculation, a 3D grid must be constructed by a transformation from the particle distribution. The complete details of this process are given in \cite{rundleetal2010}, but the basic idea is to repeatedly divide an initial cell centred on the entire disc according to some resolution criterion. The parent cell is divided once in each dimension, resulting in $2^{D}$ child cells, where $D$ is the number of dimensions, each of volume $2^{-D}$ times that of its parent. A grid resolution is imposed, such that each cell may only contain a certain amount of mass (typically $10^{-3}$ \mj is sufficient for our requirements). If the amount of mass in that cell exceeds this mass, the cell is again split once in each dimension. In this manner, child cells become parent cells and are recursively split until the mass contained in each cell is less than or equal to $10^{-3}$ \mj. 

The radiation field of the protostar is modelled using $10^9$ propagating photon packets, which undergo a random walk through the grid, experiencing scattering or absorption and re-emission until they escape the domain. Once escaped, estimates are made of the absorption rate in each cell. The dust temperatures are then determined for each cell by assuming radiative equilibrium, and the photon loop is repeated with the updated dust temperatures. This process is continued until the temperatures converge (i.e., the temperatures change very little between photon loop iterations). Continuum images are then produced for arbitrary viewing angles using the Monte-Carlo method.
 
For all of our radiative transfer results, we use typical values for a pre-main-sequence star with a central source mass of $M_*=0.6$ M$_\odot$ (i.e., the observed mass of Elias 2-27), such that $R_*=2.3$ R$_\odot$ and $T_\mathrm{eff}=3850$ K \citep{baraffeetal1998,baraffeetal2002}. The dust in our model is \cite{drainelee} silicates, with a grain size distribution given by 
\begin{equation}
n(a) \propto a^{-q}  \hspace{1cm}\mathrm{for}\hspace{1cm} a_{\mathrm{min}} < a < a_{\mathrm{max}},
\end{equation}
where $a_\mathrm{min}$ and $a_\mathrm{max}$ are the minimum and maximum grain sizes, taken to be 0.1 $\mu$m and $2000$ $\mu$m respectively, and $q=3.5$, the standard power-law exponent for the interstellar medium (ISM) \citep{grainsize}. As mentioned in section \ref{subsec:sphsetup}, reducing the opacity is equivalent to reducing the dust-to-gas mass ratio in the system. Therefore, for simulations with solar opacity, we assume the canonical dust-to-gas ratio of 0.01 everywhere in the disc. However, for simulations with opacity $0.25\times$ solar, we reduce the dust-to-gas ratio to 0.0025.

\subsection{The ALMA simulator}
\label{sec:alma}

The emission maps generated by \texttt{TORUS} are used as inputs to the ALMA simulator included in \texttt{CASA} (ver 4.7.2) \citep{CASA}, and all discs were imaged as though at a distance of 139 pc in the $\rho$-Ophiuchus region \citep{mamajek2008}, at the same sky position of Elias 2-27. So that we are able to directly compare with the original observation of Elias 2-27, we simulated the observations with the antenna configuration originally used by \citet{perezetal2016} in cycle 2 (selected via the antennalist parameter in \texttt{CASA}, and in this particular case is named alma.cycle2.6.cfg). We used the \texttt{restoring beam} function, which allows the user to replicate the exact beam used in a given observation.

In this work, we use the same position angle (117.3\textdegree) and inclination (55.8\textdegree{}) for our synthetic images as reported by \citet{perezetal2016} in the original observation. We do not consider the effect of rotating the system or varying the inclination, but we note that doing so may result in spiral arms being hidden or revealed to the observer as optical depth along the line of sight changes.

The total time on each of the sources was 12 minutes, as in the original observation, centred on 230 GHz with a bandwidth of 6.8 GHz. Visibilities were corrupted using the Atmospheric Transmission at Microwaves (\texttt{ATM}) code \citep{ATMcode}, with a precipitable water vapour value of 2.784 mm. The total observation time is low when compared to some recent investigations into the detectability of gravitational instabilities in protostellar discs using ALMA (see e.g. \citealt{dipierro2015}), however, positive results have been obtained by \citet{meruetal2017} using identical on-source times to this work.

\subsection{Unsharp image masking}
\label{sec:unsharp}
The purpose of unsharp image masking \citep{unsharp} is to reduce the overall range in flux of the image, without reducing the range of individual details. This enhances fainter features that are otherwise washed out in an image with a high flux range. 

We perform the unsharp image masking on the synthetic observations described in section \ref{sec:alma} exactly as described in \citet{perezetal2016}, by subtracting a smoothed and scaled version of the data from itself. We smoothed the original data with a 2D Gaussian of 0.33$^{\prime\prime}\times 0.33^{\prime\prime}$, before scaling it by a factor of 0.87 and subtracting it from the original data.

\begin{figure}
\centering
\includegraphics[width=0.5\textwidth]{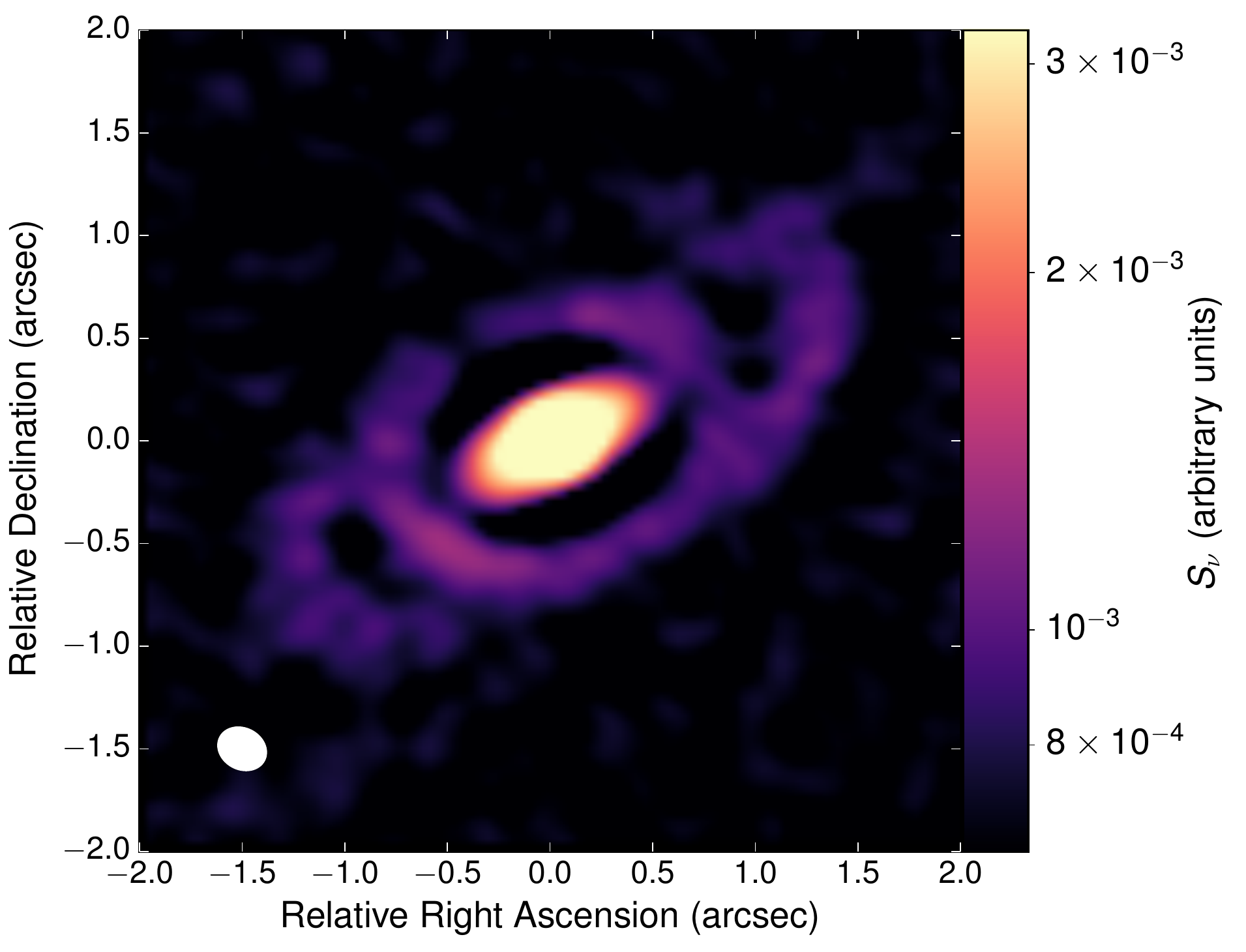}
\caption{The original observation of the thermal dust emission of the Elias 2-27 system, processed with the unsharped image masking technique outlined in section \ref{sec:unsharp}.\label{fig:unsharp}}
\end{figure}

To be confident that our image analysis matched that of \cite{perezetal2016}, we applied the same technique to the original continuum image (available from the ALMA archive), and our result is shown in Figure \ref{fig:unsharp}. As in the original work, a deep cavity appears on the image minor axis due to inclination effects, and the central core is not completely removed.

%
\section{Results}
\label{sec:results}
Table \ref{tab:simulations} shows the results of the SPH simulations. Out of a total of 17 models,  only 6 developed spiral structure in the SPH simulations and did not fragment. In all cases, a minimum temperature floor was used to allow a quasi-steady, non-fragmenting state to develop. 

The 6 discs that showed spiral structure were used as input to the \texttt{TORUS} code as described in section \ref{sec:torus}, before being observed with the ALMA simulator as outlined in section \ref{sec:alma}. The unsharped masking technique described in section \ref{sec:unsharp} was applied to the ALMA simulator output.

\begin{figure*}
  \centering
  \begin{tabular}{ccc}
    0.25 $\times$ Solar &      1.0$\times$ Solar &    1.0$\times$ Solar \\
    $q=0.325$                &       $q=0.325$            &    $q=0.325$   \\
    $T_\mathrm{min}=5$ K &  $T_\mathrm{min}=5$ K &  $T_\mathrm{min}=10$ K \\
\includegraphics[width=0.25\textwidth,angle=90]{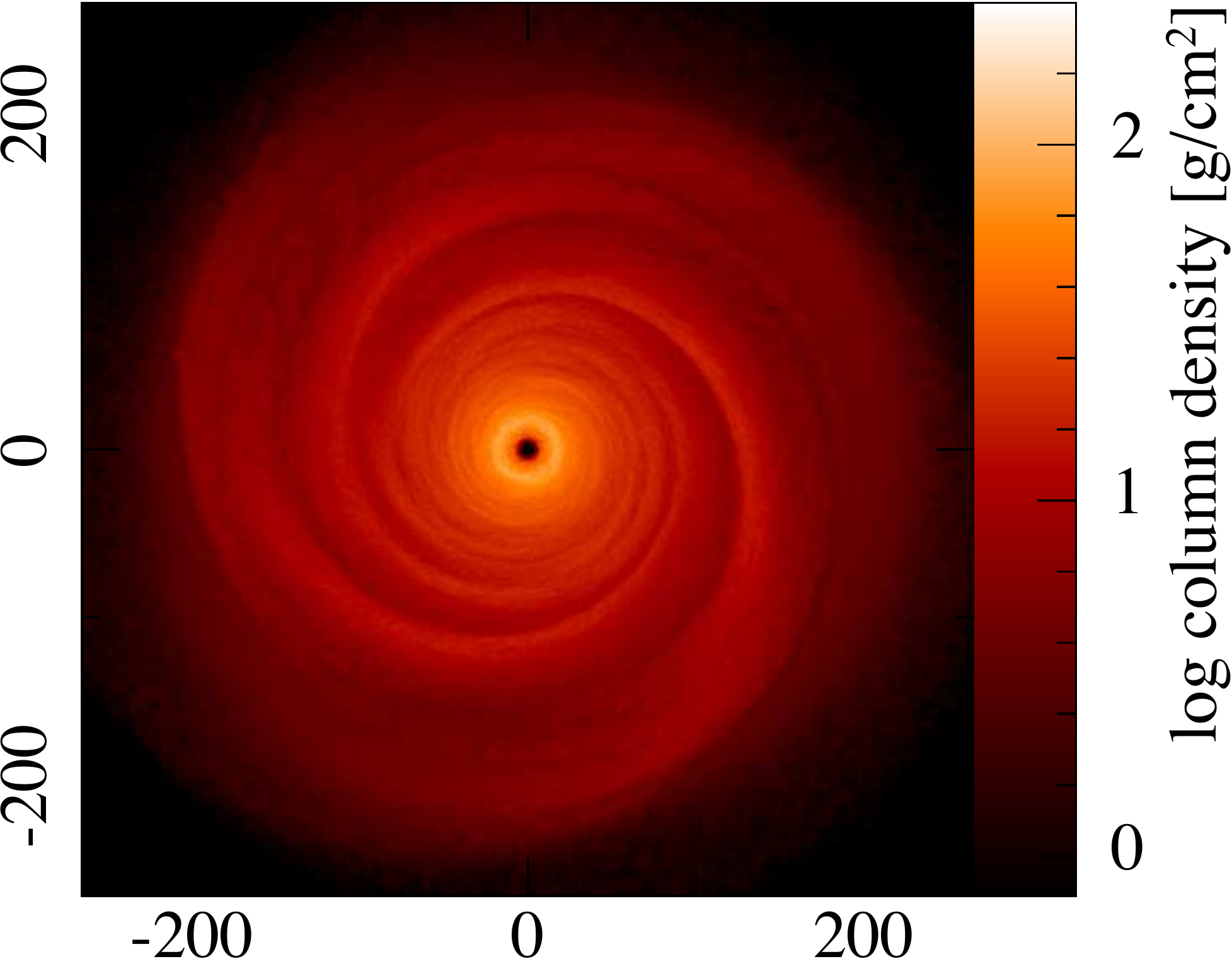} &\includegraphics[width=0.25\textwidth,angle=90]{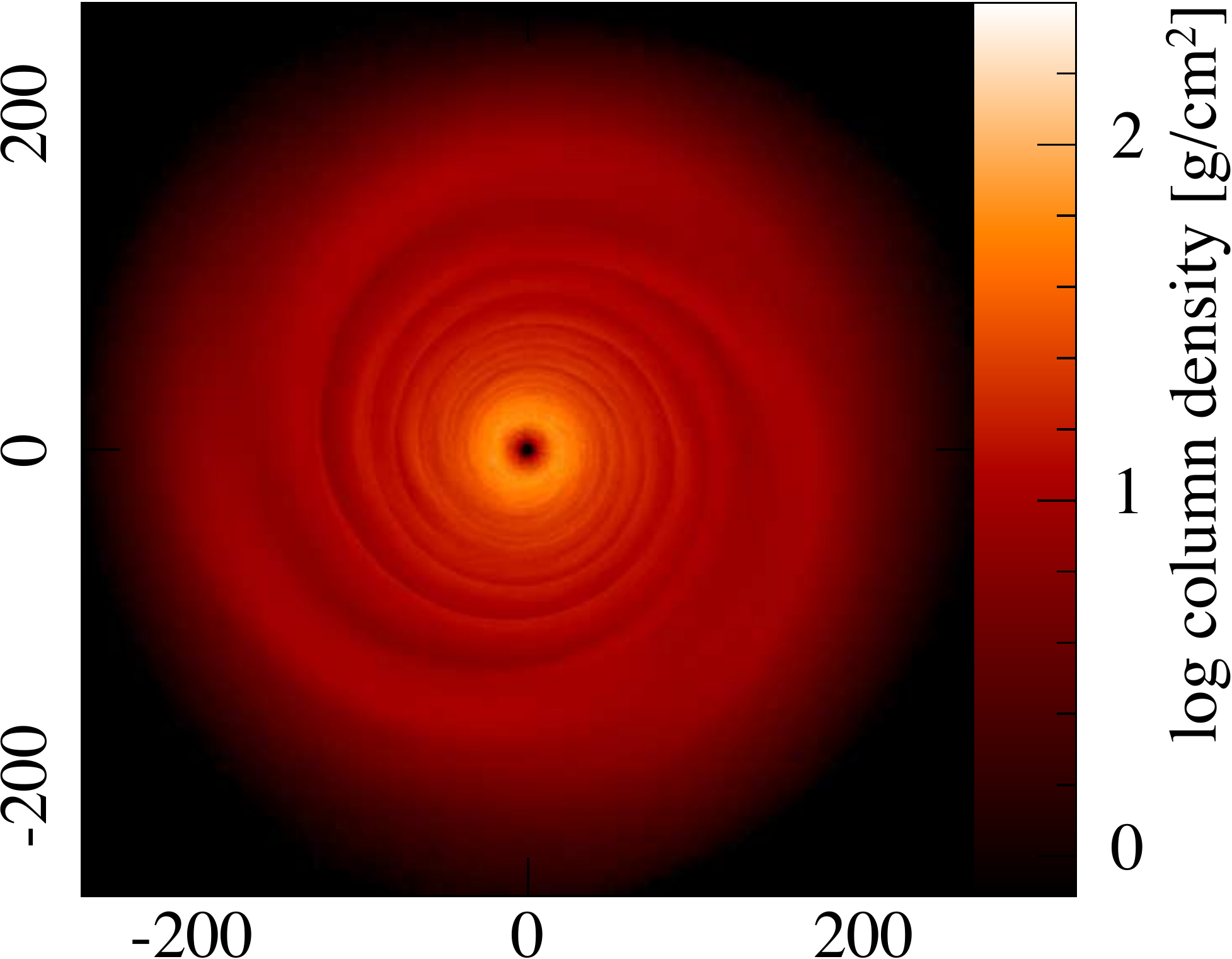} & \includegraphics[width=0.25\textwidth,angle=90]{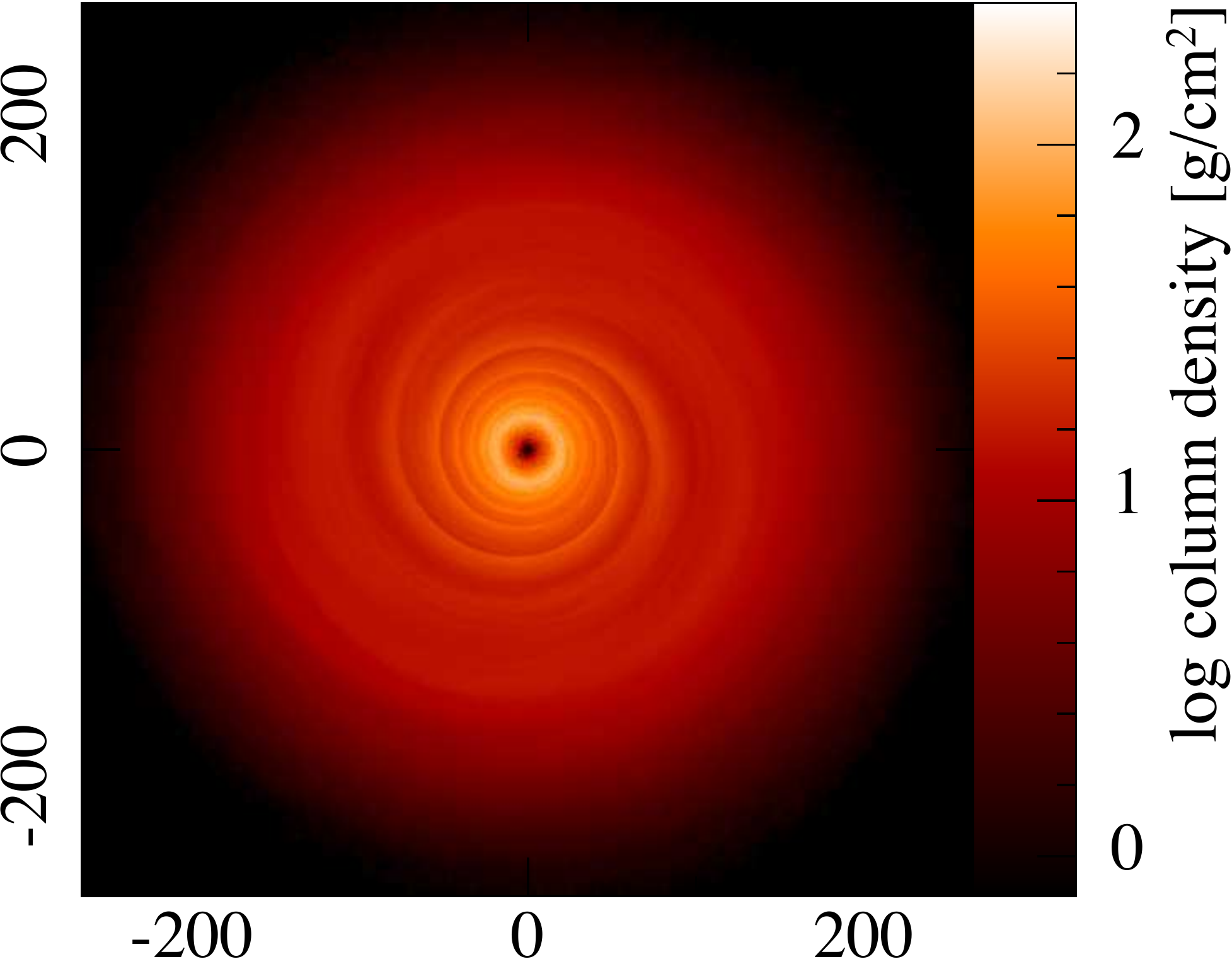} \\
\includegraphics[width=0.33\textwidth]{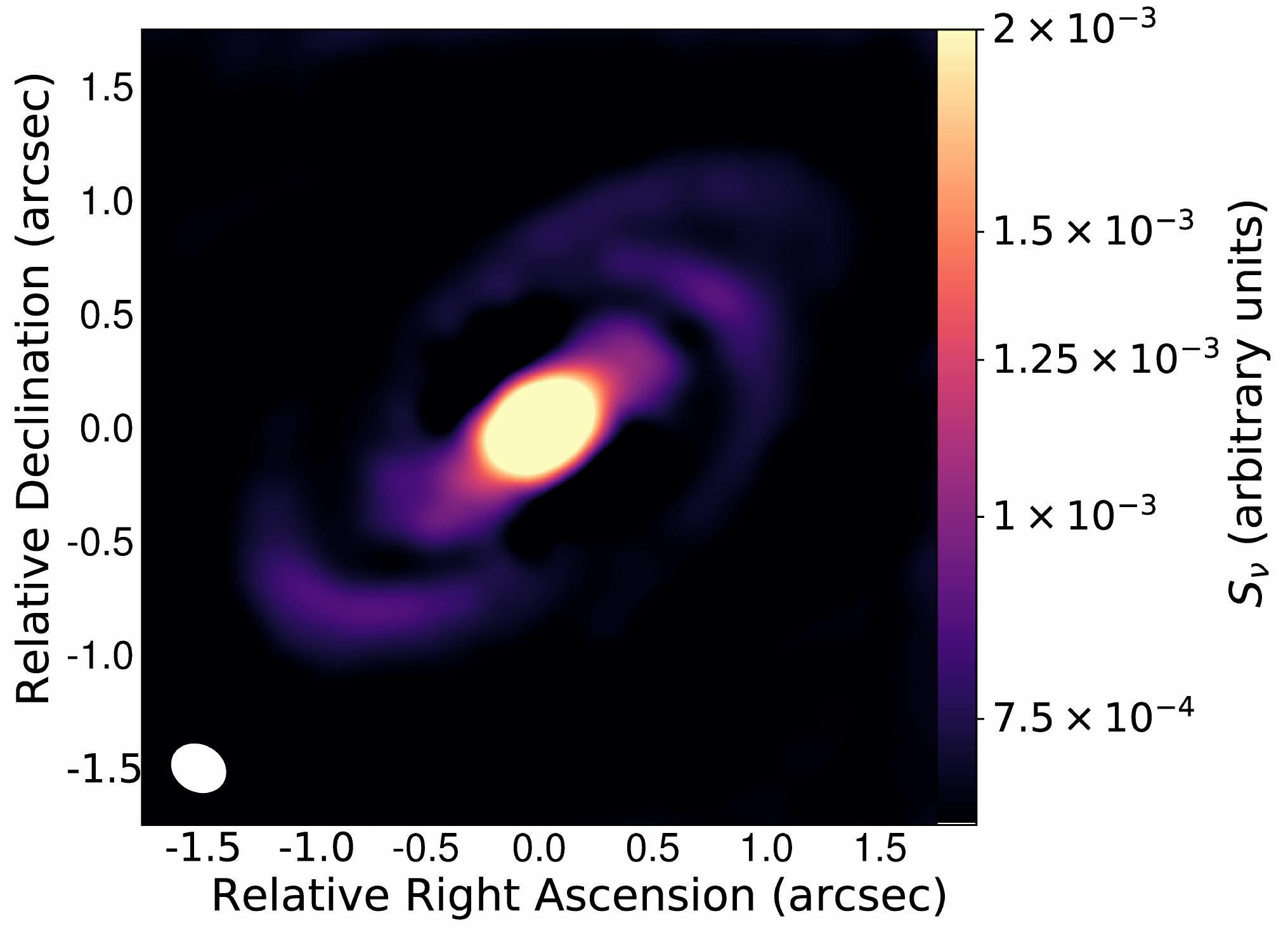} & \includegraphics[width=0.33\textwidth]{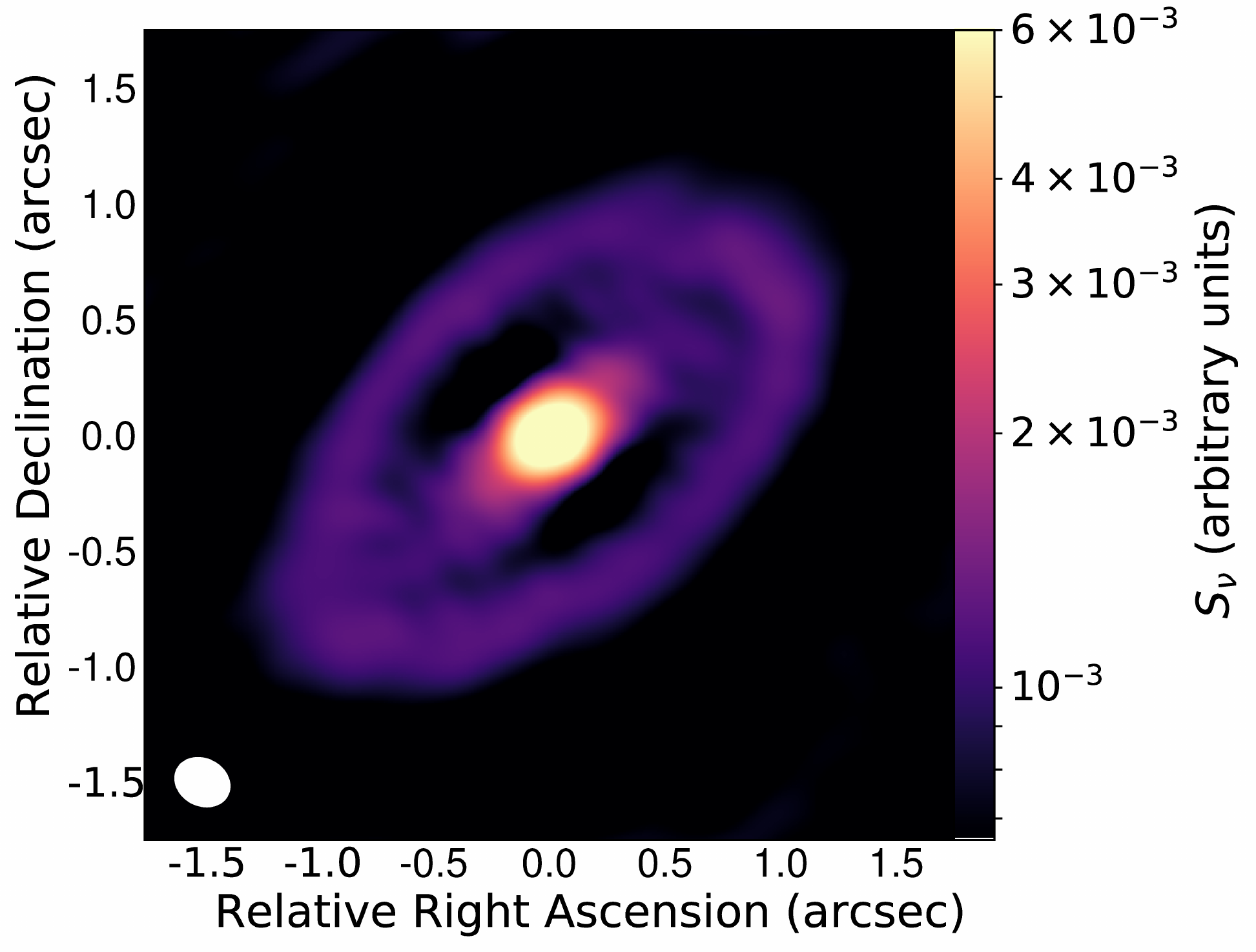} & \includegraphics[width=0.33\textwidth]{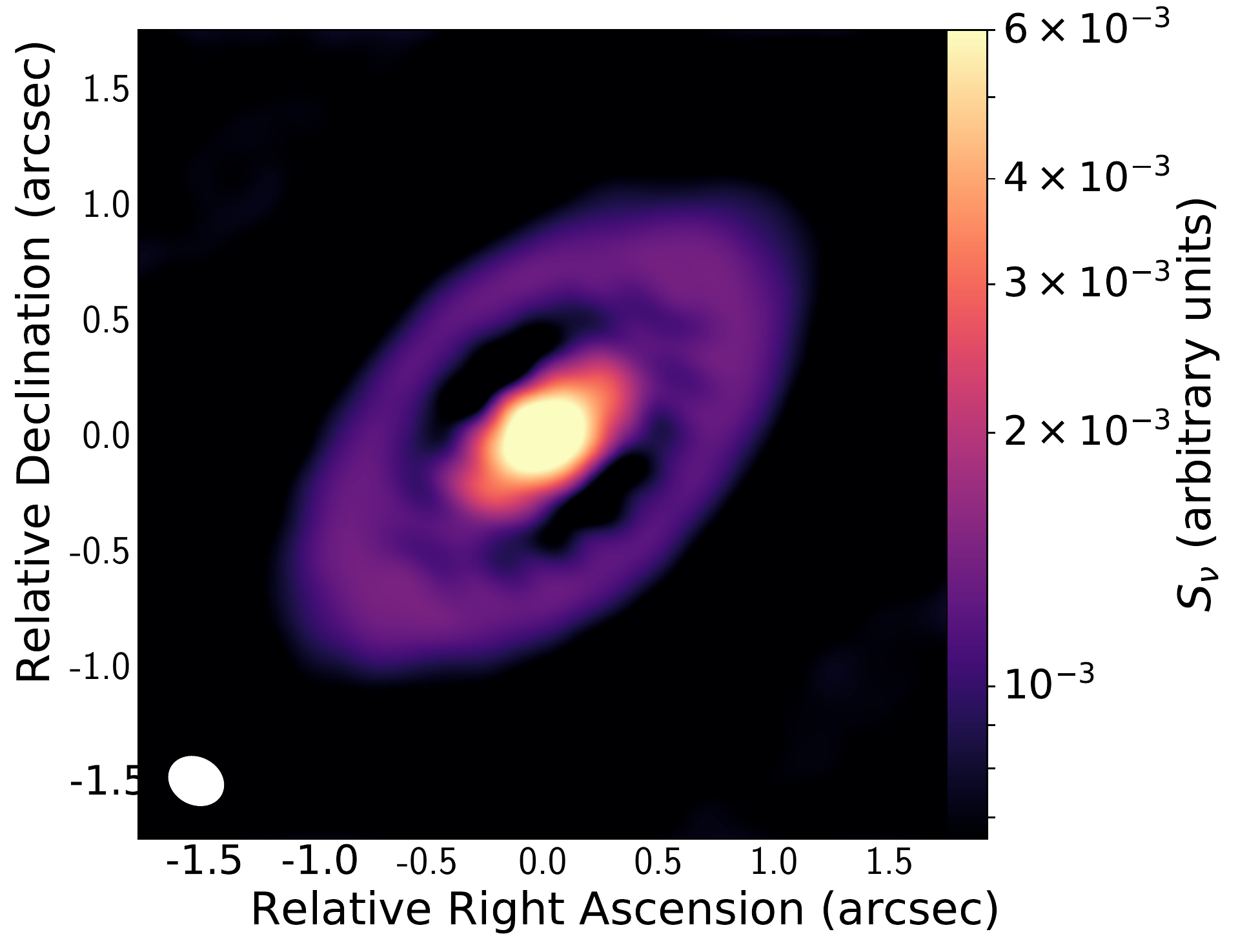}
  \end{tabular}
  \caption{All discs shown in this Figure have the same disc-to-star mass ratio of $q=0.325$. Top row shows log column density plots of the SPH simulation results, bottom row shows how the discs would be viewed with ALMA, after using the image enhancing technique described in section \ref{sec:unsharp}. Leftmost column has metallicity (and therefore, we assume, opacity) $0.25\times$ that of solar, with a minimum temperature of $T=5$ K. Center column has solar metallicity with minimum temperature of $T=5$ K, and rightmost column has solar metallicity with a minimum temperature of $T=10$ K. As can be seen from the synthetic images displayed on the bottom row, only the disc with the lowest opacity has a detectable spiral structure. This is due to the lower opacity resulting in more efficient cooling, and discs with more efficient cooling have larger spiral amplitudes, and, therefore, typically have larger density contrast between the arm and inter-arm region. The overall flux is lower in lower opacity disc due to less dust being present in the system.\label{fig:lowmassdiscs}}
\end{figure*}

\begin{figure*}
  \centering
  \begin{tabular}{ccc}
    0.25 $\times$ Solar        &      1.0$\times$ Solar      &    1.0$\times$ Solar \\
    $q=0.4$                           &       $q=0.4$                      &    $q=0.5$   \\
    $T_\mathrm{min}=10$ K &  $T_\mathrm{min}=10$ K &  $T_\mathrm{min}=10$ K \\
    \includegraphics[width=0.25\textwidth,angle=90]{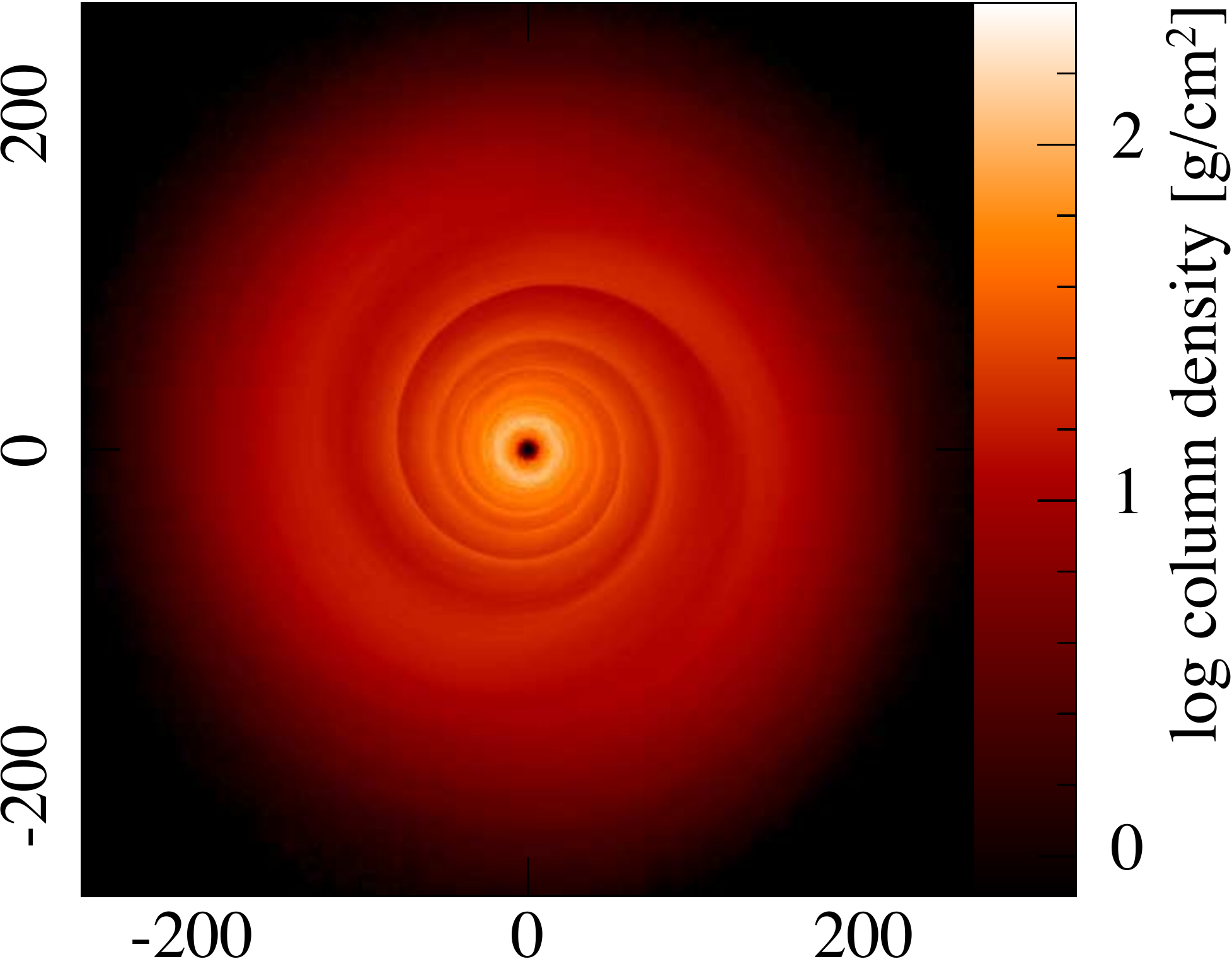} & \includegraphics[width=0.25\textwidth,angle=90]{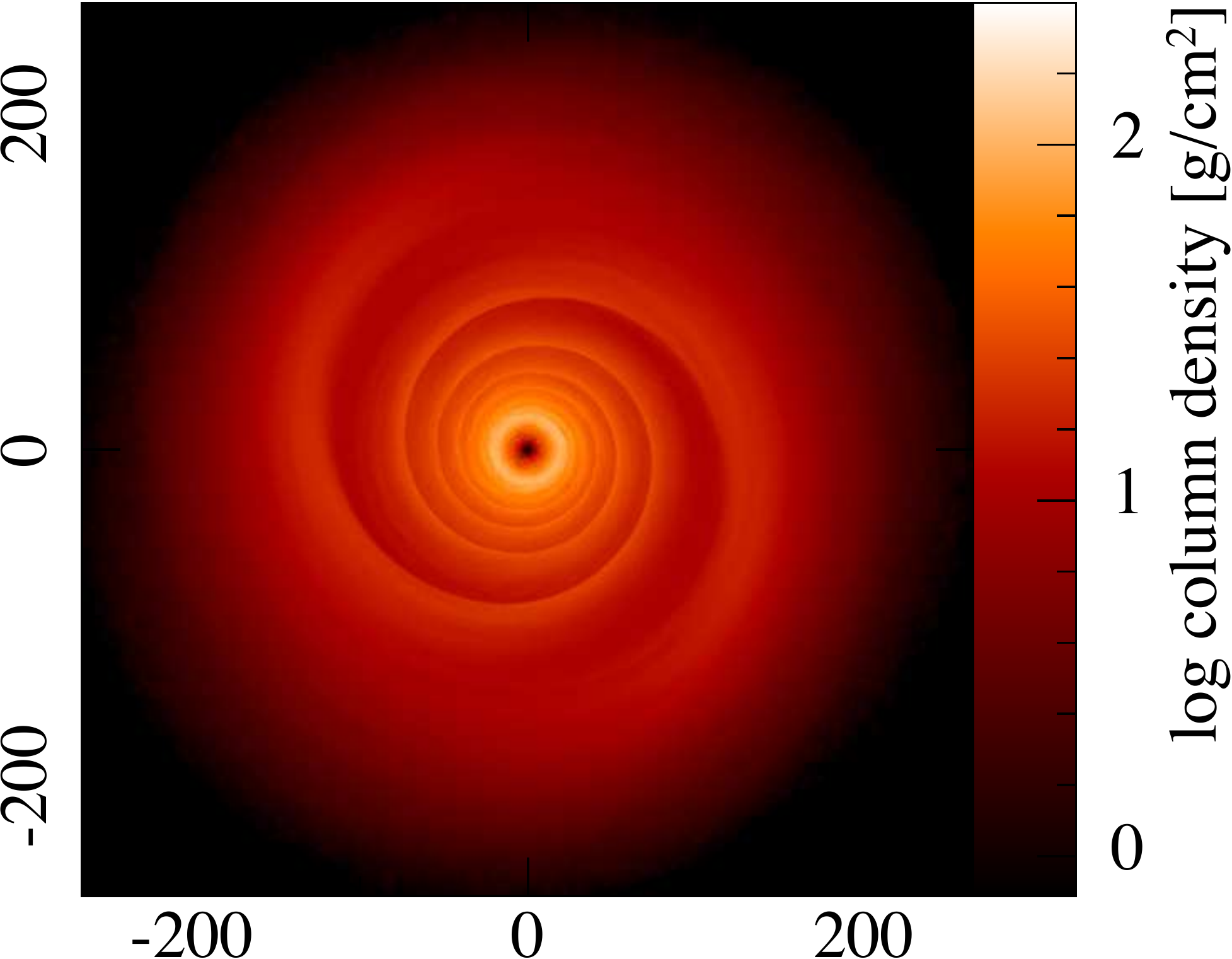} & \includegraphics[width=0.25\textwidth,angle=90]{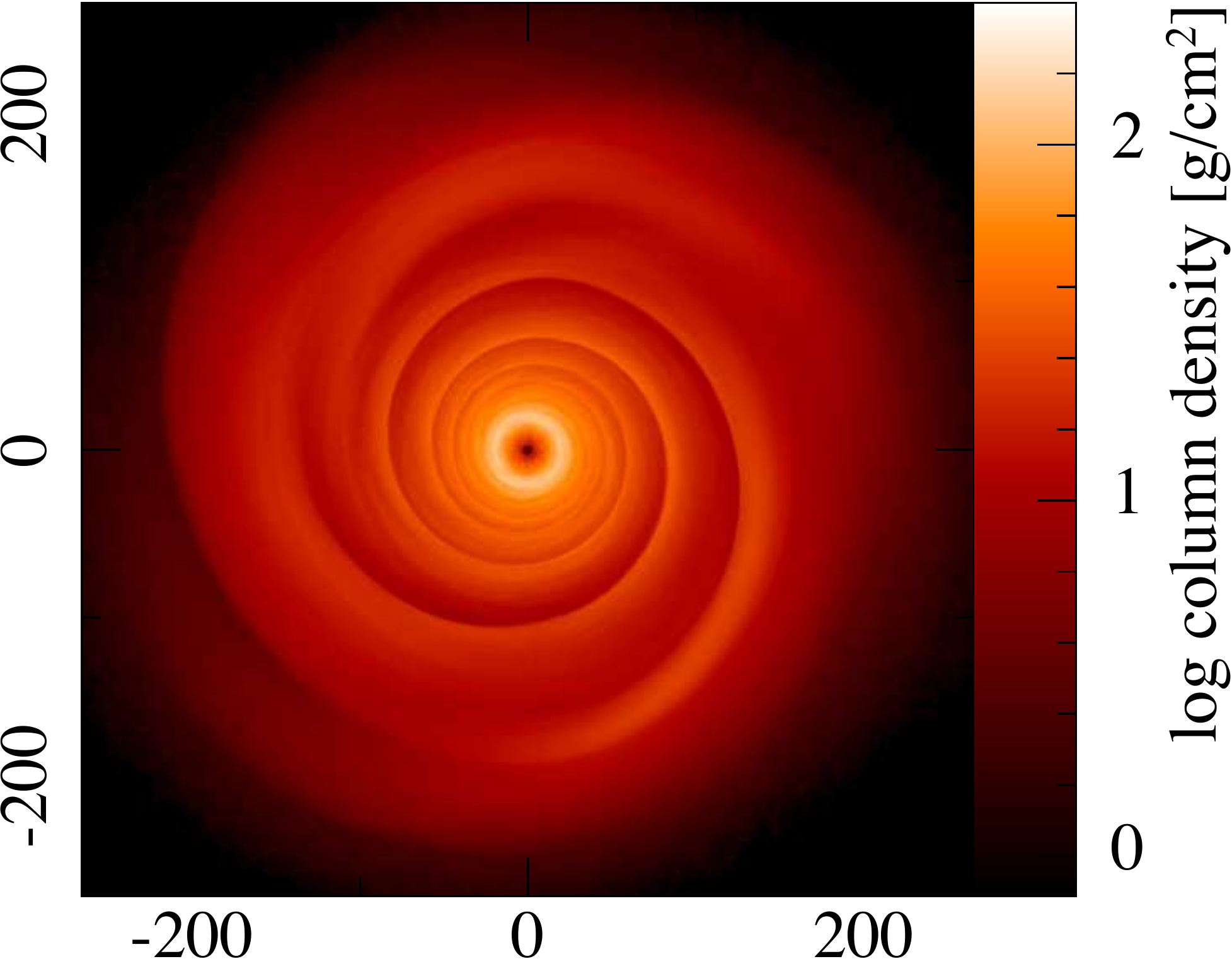} \\
    \includegraphics[width=0.33\textwidth]{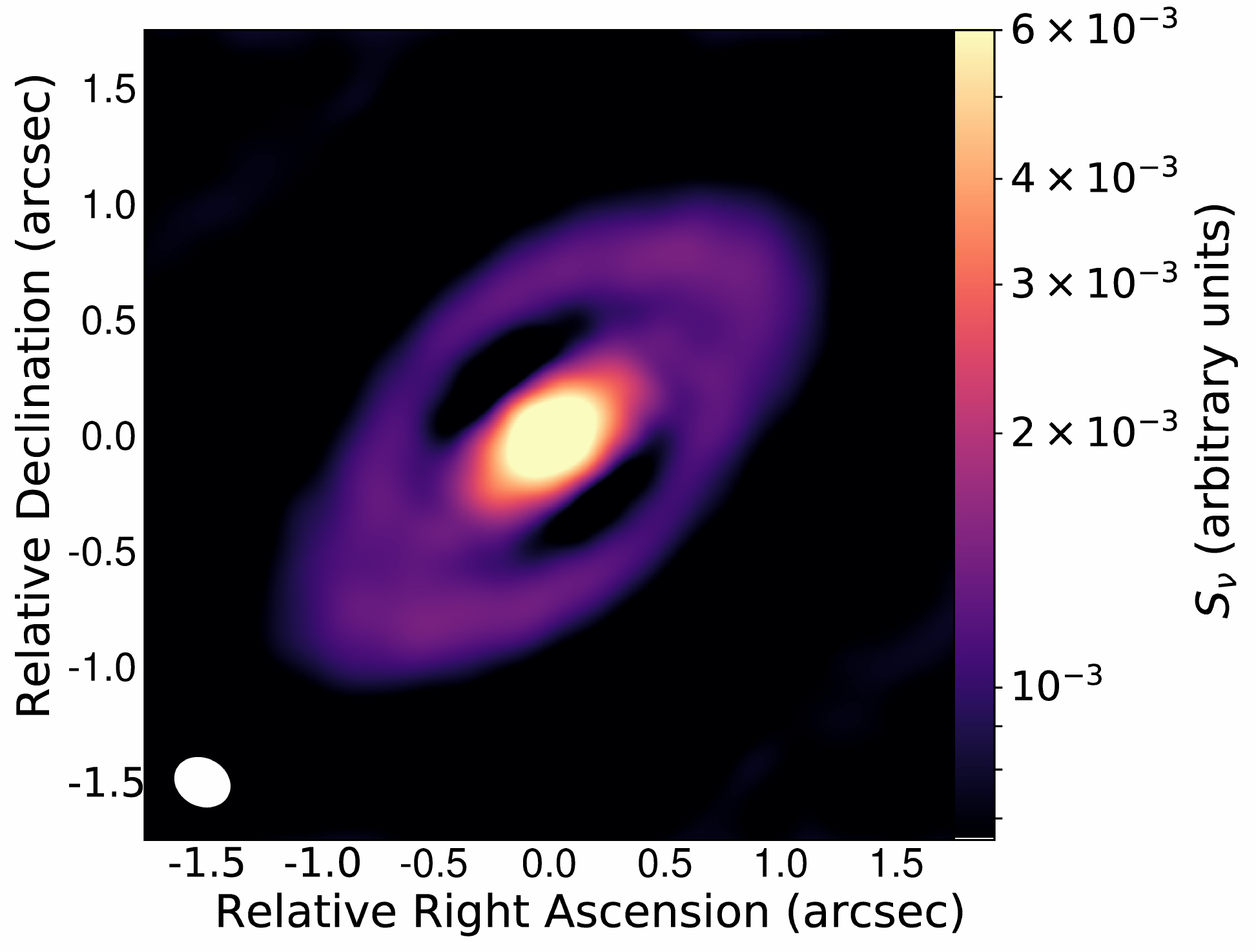} & \includegraphics[width=0.33\textwidth]{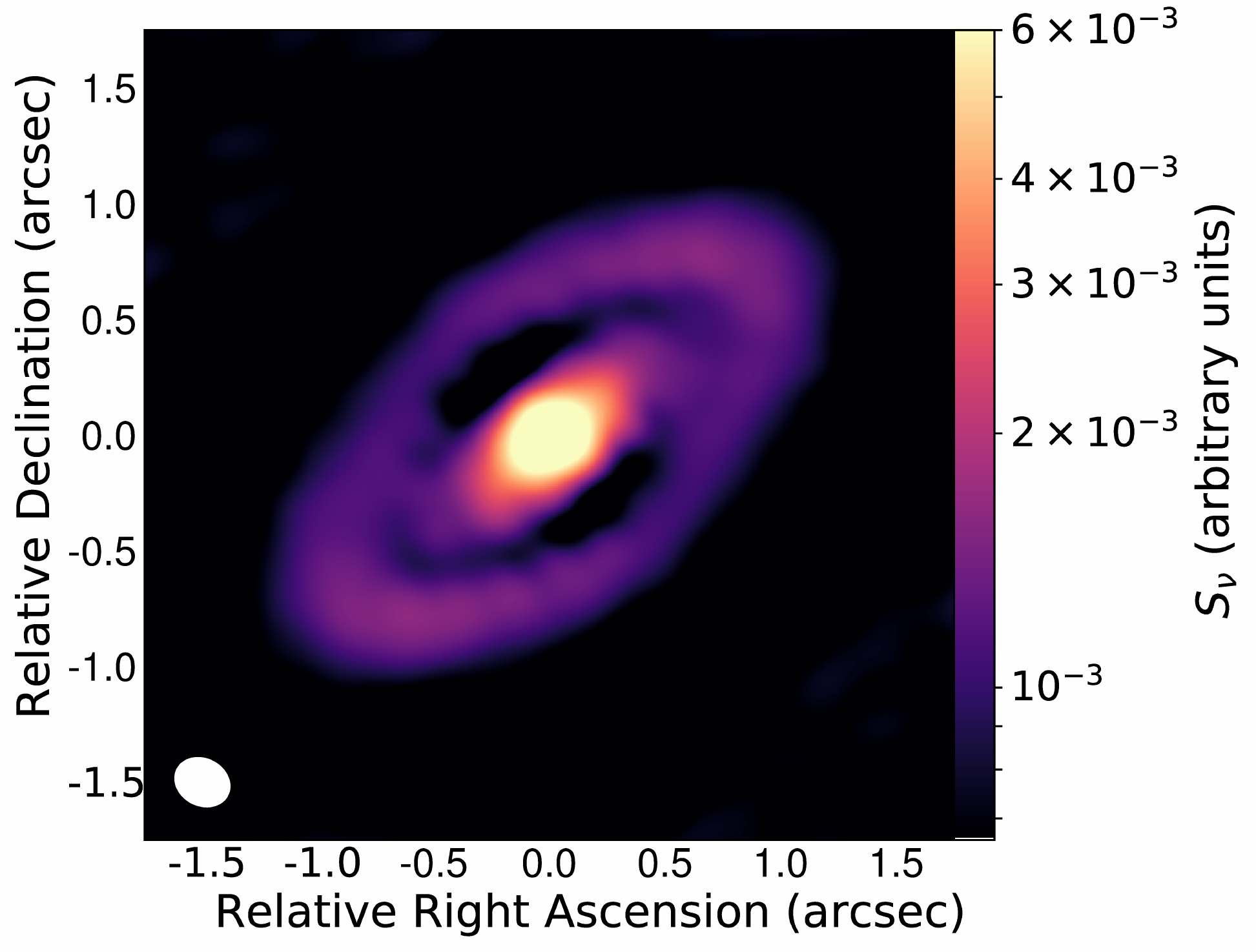} & \includegraphics[width=0.33\textwidth]{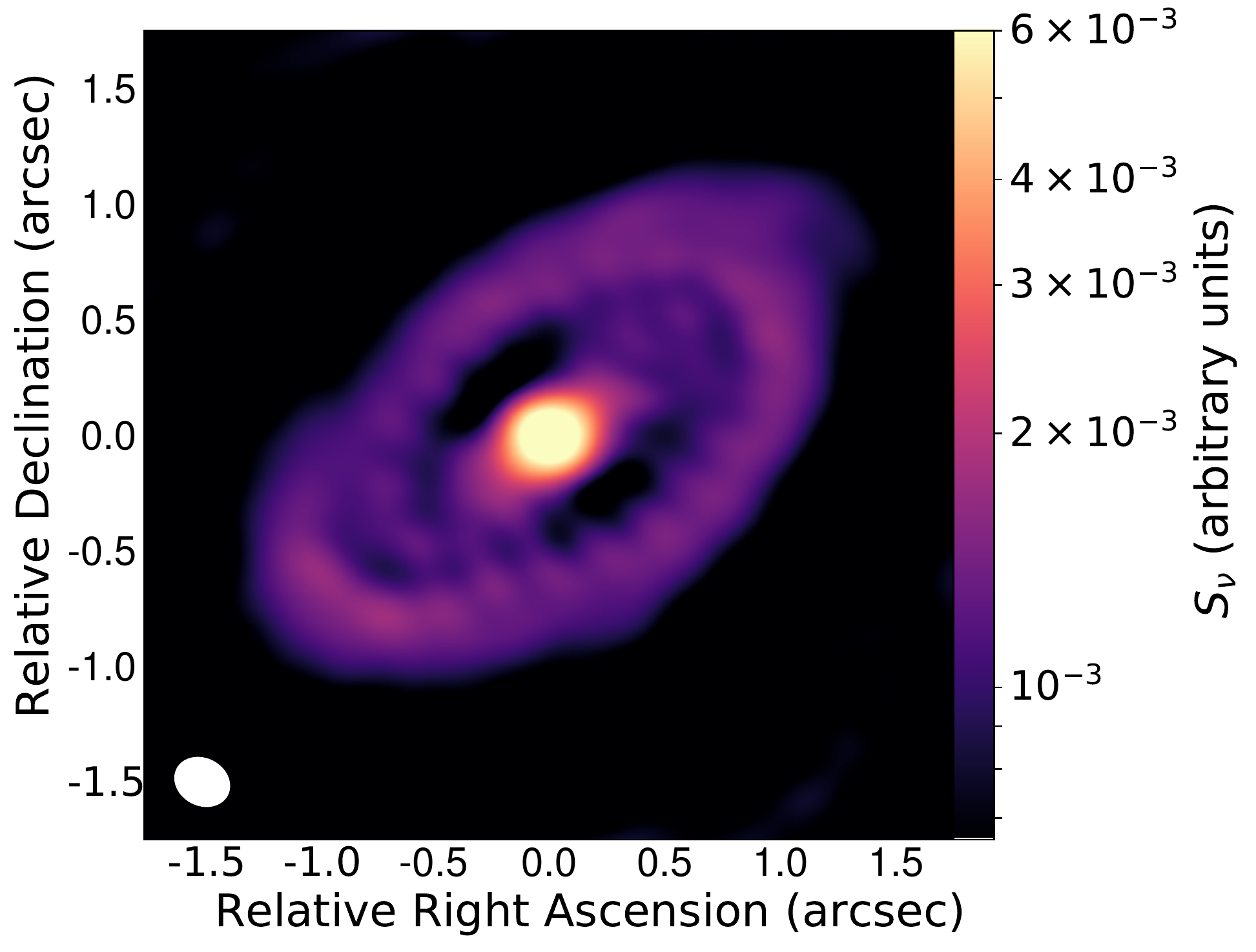} 
    
  \end{tabular}
  \caption{Left and center columns in this figure show discs with a disc-to-star mass ratio of $q=0.4$, while the right column shows a disc with disc-to star mass ratio of $q=0.5$. The top row shows log column density plots of SPH simulation results, bottom row shows how the discs would be viewed with ALMA, after using the image enhancing technique described in section \ref{sec:unsharp}. Leftmost column has metallicity (and therefore, opacity) $0.25\times$ that of solar, with a minimum temperature of $T=10$ K. Center column has solar metallicity with a minimum temperature of $T=10$ K, and rightmost column has solar metallicity with a minimum temperature of $T=10$ K. Synthetic images of each disc are displayed on the bottom row. In all cases, a minimum  temperature of 10 K is enough to reduce the spiral amplitude to below a level detectable by ALMA with an integration time of 12 minutes on source, used in the original observation of Elias 2-27 by \citet{perezetal2016}. \label{fig:highmassdiscs}}
\end{figure*}

The results are shown in Figures \ref{fig:lowmassdiscs} and \ref{fig:highmassdiscs}. Figure \ref{fig:lowmassdiscs} shows three discs with disc-to-star mass ratio of $q=0.325$, left column has metallicity $0.25\times$ solar, center and right have solar metallicity, left and center columns have minimum temperature $T_{\mathrm{min}}=5$ K, right column has $T_{\mathrm{min}}=10$ K. The top row shows column density renderings of the SPH discs, while the bottom row shows the synthesised ALMA images of these discs at 230 GHz (1.3 mm).

Only in the low metallicity (and therefore low opacity) case (leftmost column) is spiral structure visible. When opacity is raised to solar, resulting in a decreased cooling efficiency and therefore weaker spiral amplitudes, the spiral structure is not detectable in the final image, producing instead ring-like structures. We note that the total flux is reduced in the systems with low opacity, this is due dust mass being reduced by a factor of 4. However, we use arbitrary units in this work, and are concerned primarily with contrast ratios rather than total values.

Similar results are found in Figure \ref{fig:highmassdiscs}, where the left and center columns show discs with a disc-to-star mass ratio of $q=0.4$, and the rightmost column shows $q=0.5$. The left column has metallicity $0.25\times$ that of solar, while the centre and right columns have solar metallicity. In all of these cases, where $q\geq 0.4$ the disc appears to have a ring-like, rather than spiral-like, morphology. In all cases shown in Figure \ref{fig:highmassdiscs} , the minimum temperature required to prevent fragmentation is $T_{\mathrm{min}}=10$ K. Star forming regions do not, typically, have temperatures below 10 K. As the cloud cores collapse, the gas becomes thermally coupled to the dust and maintains an almost constant temperature of $\sim 10$ K over a large range of densities \citep{hayashinakano1965,tohline1982, larson1985,masunagainutsuka2000}. This is backed by observations of even the coldest star formation regions (e.g., Table 2 of \citealt{battersbyetal2014}), and simulations of cores with young protostars maintaining temperatures of $15-30$ K \citep{stamatellosetal2005a,stamatellosetal2005b}.  

This brings us to our first conclusion. Even if spirals generated by gravitational instability are present in a protostellar disc, it may be difficult to detect them unless they exist in the narrow region of parameter space that favours their detection. In this case, we apply the ideas discussed in \citet{halletal2016}; that for a protostellar disc to have sufficiently large spiral amplitudes so as to be detected, but not so large as to cause the disc to fragment, the mass and accretion rate of the disc can only have a narrow range of values. This parameter space can be broadened somewhat by partially supporting the disc against fragmentation by bathing the system in some external irradiation, but temperatures that are too high will reduce the amplitude of the spirals below a detectable level. 

%
%

\subsection{Spiral amplification results}
\begin{figure*}
  \centering
  \begin{tabular}{ccc}
    0.25 $\times$ Solar &      1.0$\times$ Solar &    1.0$\times$ Solar \\
    $q=0.325$                &       $q=0.325$            &    $q=0.325$   \\
    $T_\mathrm{min}=5$ K &  $T_\mathrm{min}=5$ K &  $T_\mathrm{min}=10$ K \\
    \includegraphics[width=0.25\textwidth,angle=90]{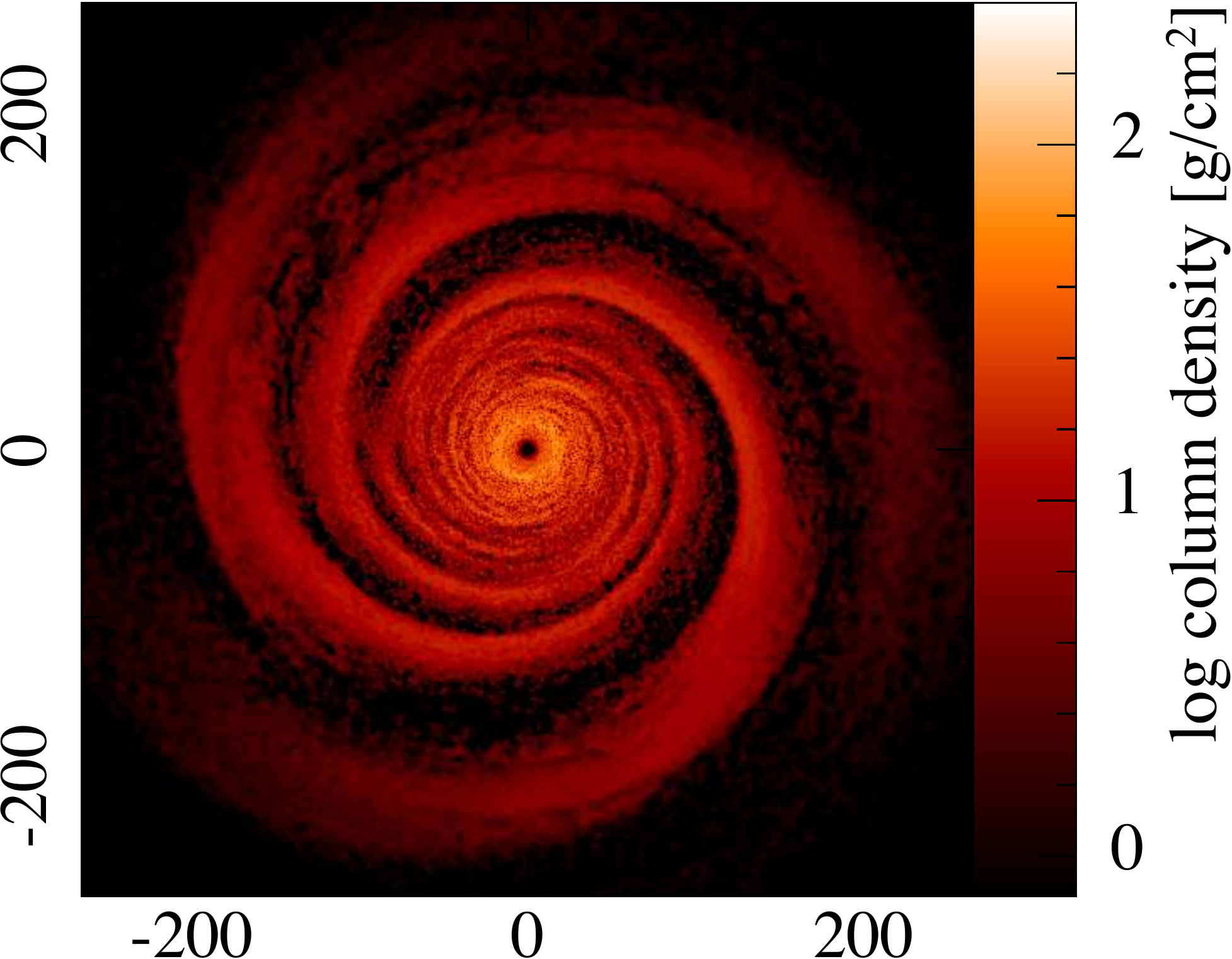} &\includegraphics[width=0.25\textwidth,angle=90]{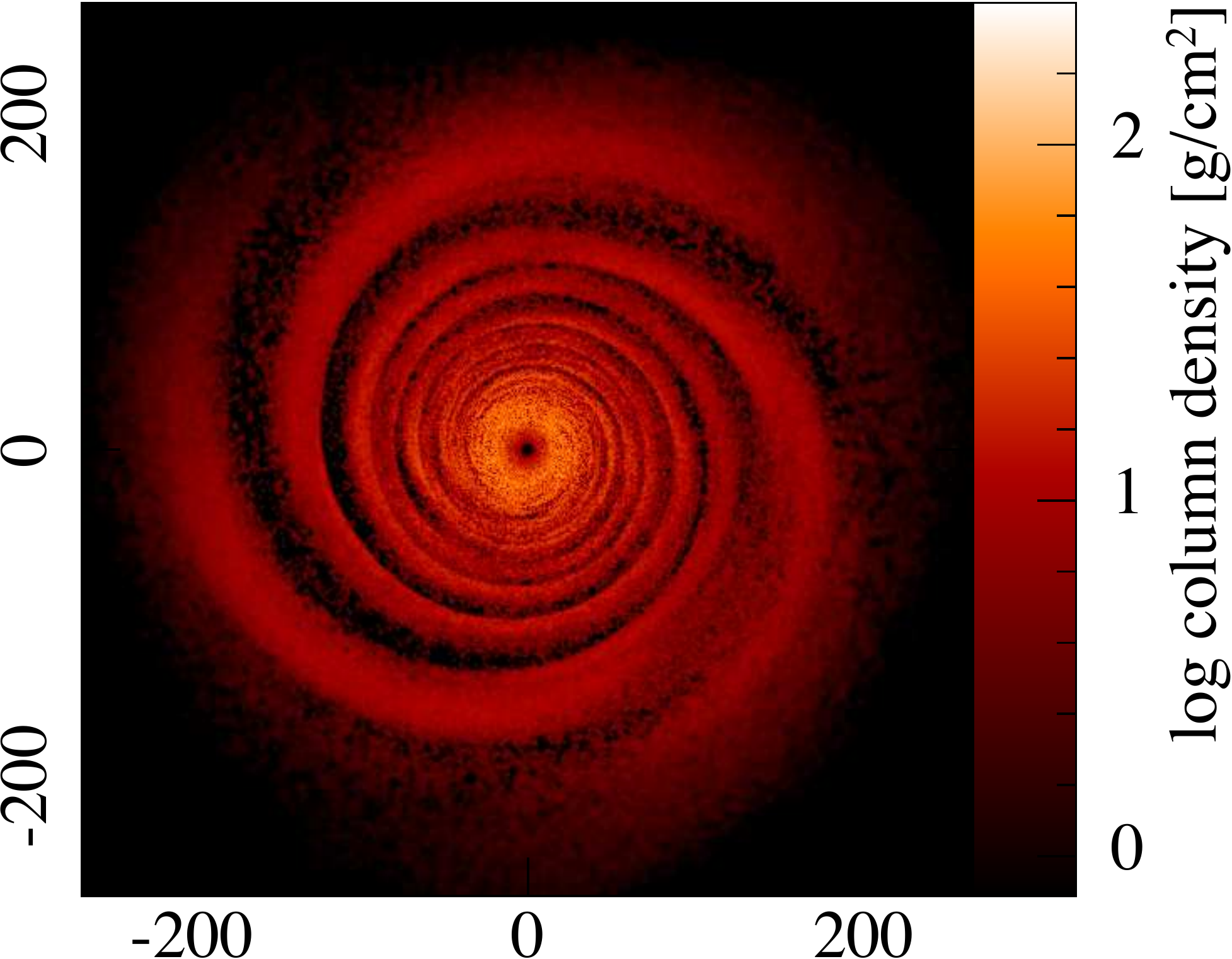} & \includegraphics[width=0.25\textwidth,angle=90]{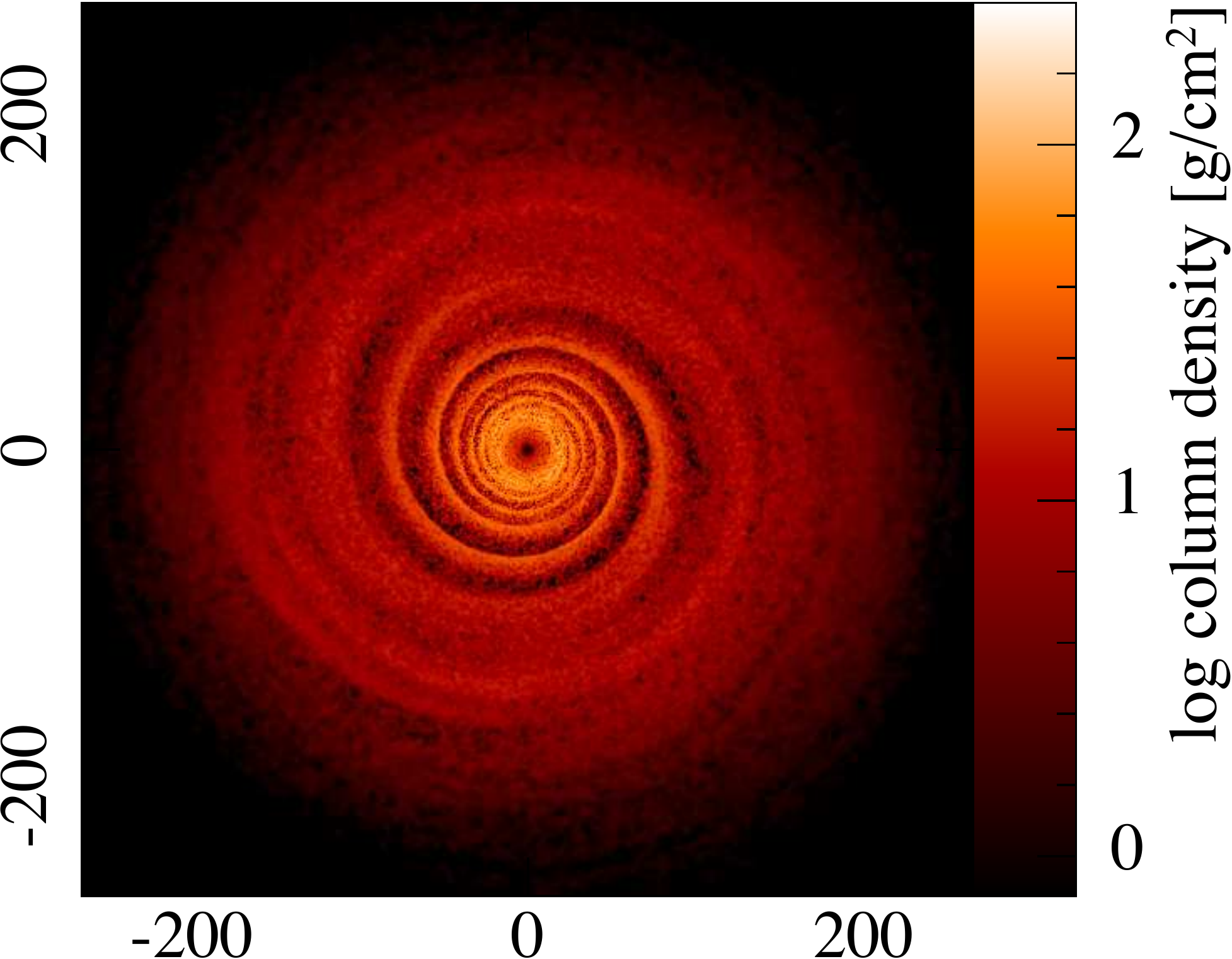} \\
       \includegraphics[width=0.33\textwidth]{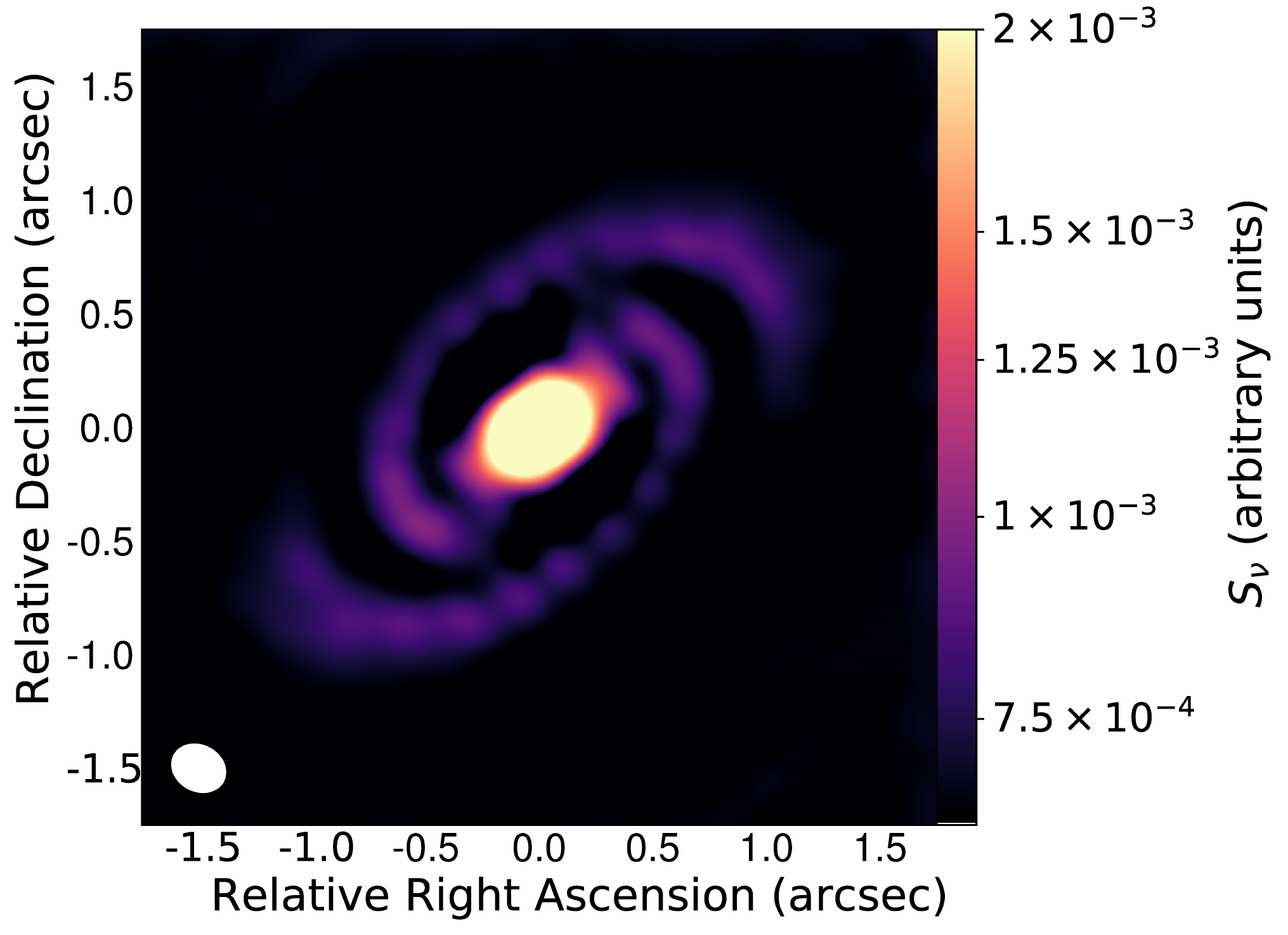} &\includegraphics[width=0.33\textwidth]{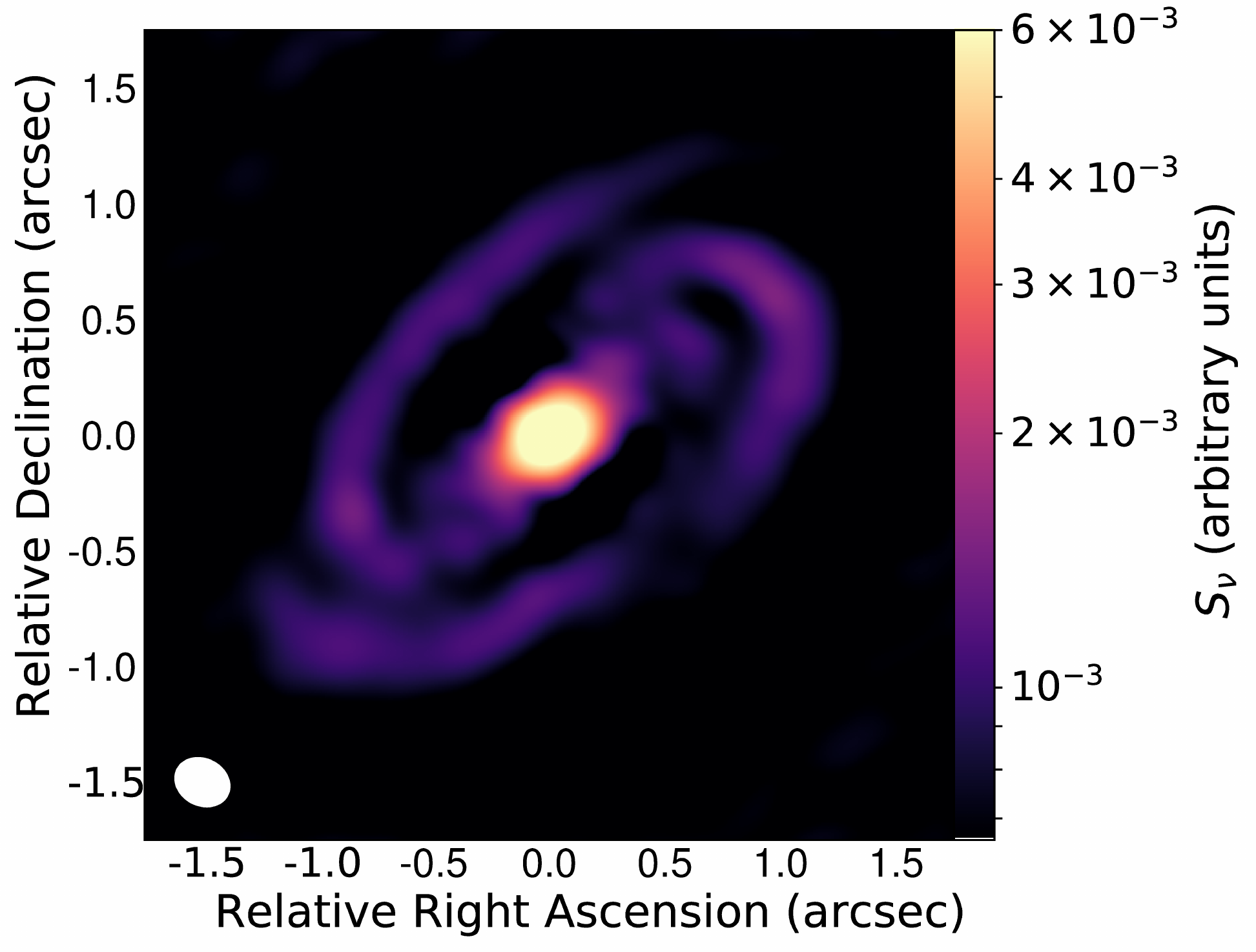} & \includegraphics[width=0.33\textwidth]{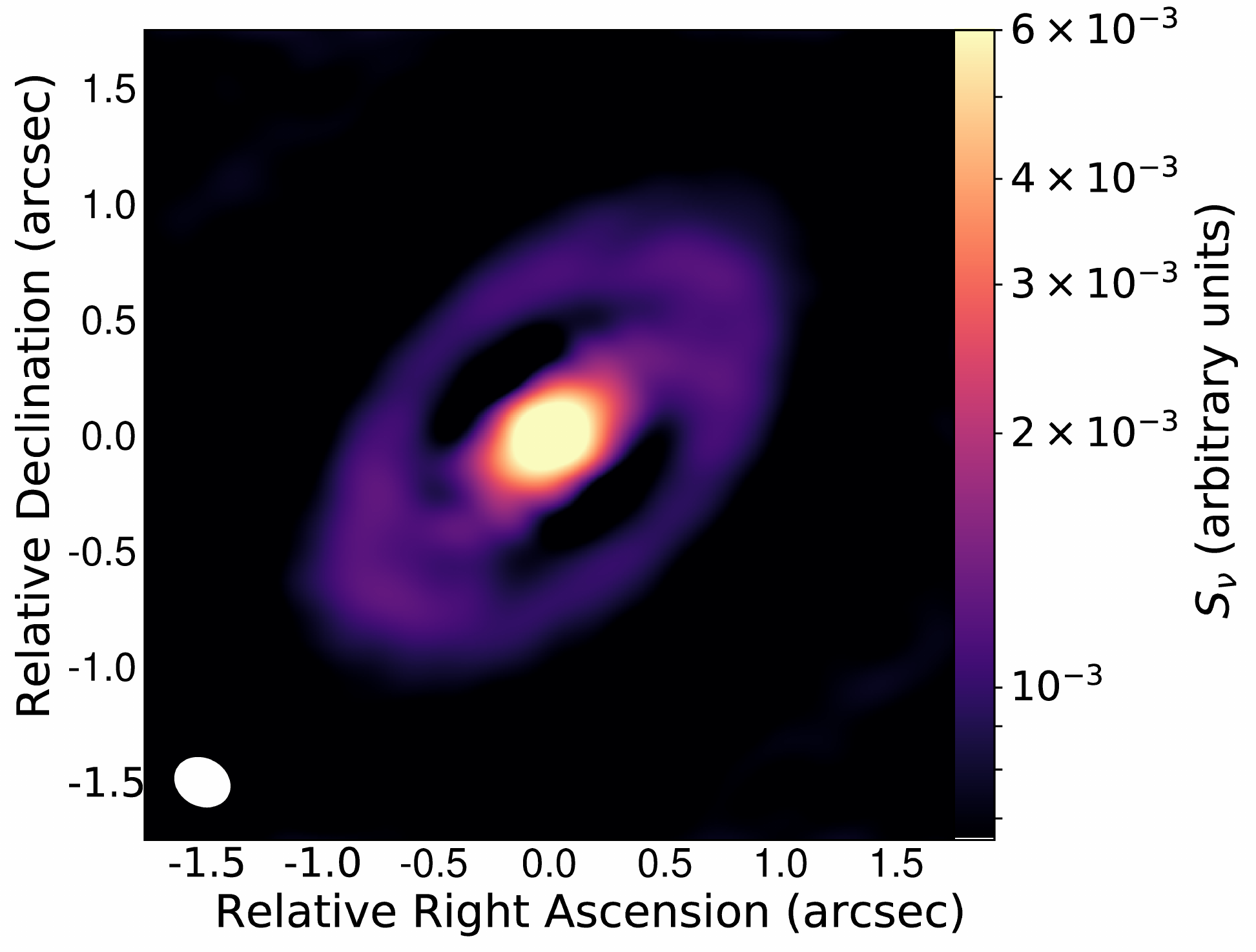} 
  \end{tabular}
  \caption{This Figure shows the same discs as Figure \ref{fig:lowmassdiscs}, but with the spirals amplified by a factor of 1.15, to account for the concentration of dust grains in the spiral arms. In the left and center column, non-axisymmetric structure is now clearly visible in the simulated ALMA image. However, in the rightmost column, even amplification does not produce detectable structure in the outer part of the disc. A minimum temperature of 10 K is enough to remove almost all spiral structure in the original disc, leaving very little structure to amplify.\label{fig:lowmassamp}}
  \end{figure*}

\begin{figure*}
  \centering
  \begin{tabular}{ccc}
    0.25 $\times$ Solar        &      1.0$\times$ Solar      &    1.0$\times$ Solar\\
    $q=0.4$                           &       $q=0.4$                      &    $q=0.5$   \\
    $T_\mathrm{min}=10$ K &  $T_\mathrm{min}=10$ K &  $T_\mathrm{min}=10$ K \\
    \includegraphics[width=0.25\textwidth,angle=90]{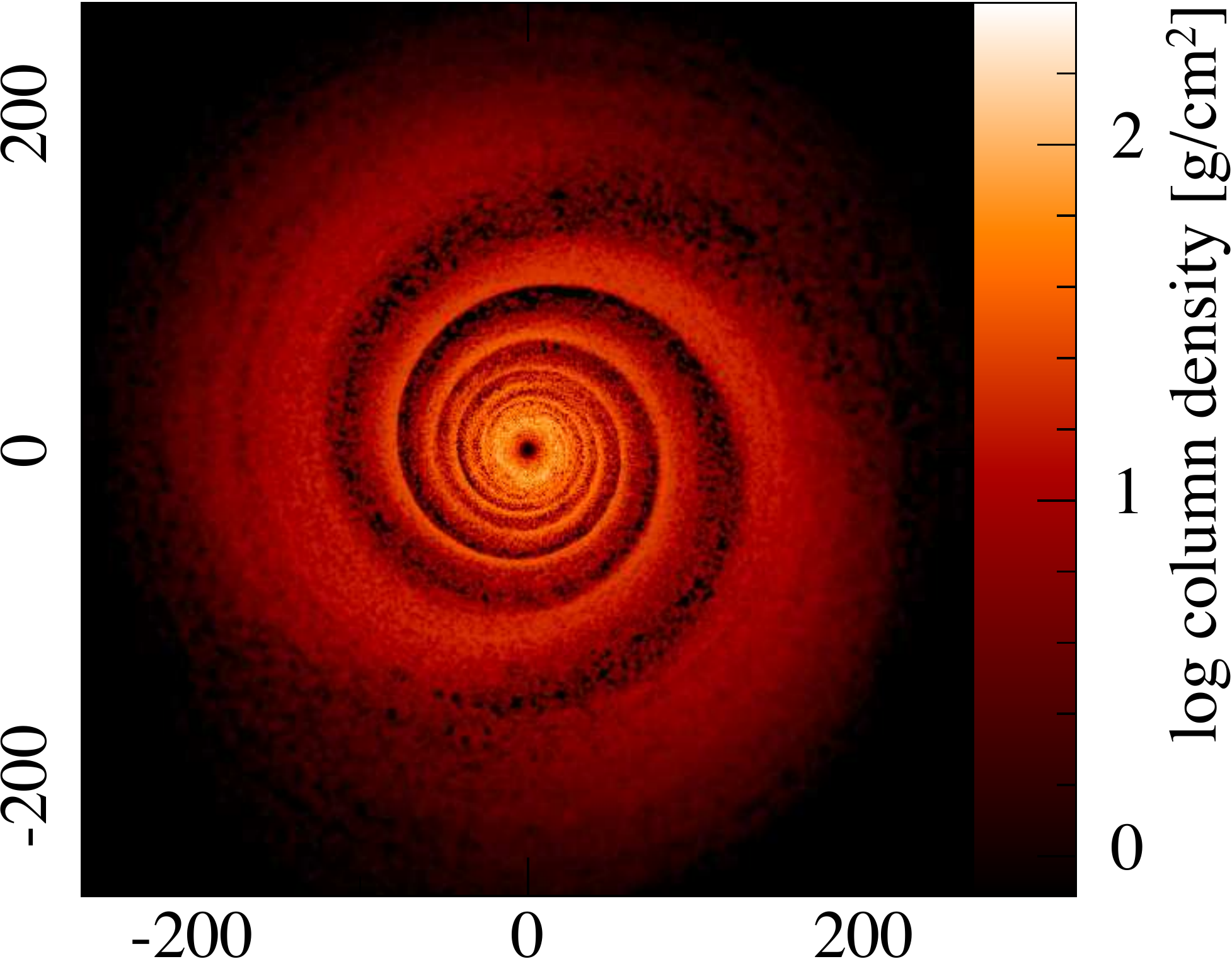} &    \includegraphics[width=0.25\textwidth,angle=90]{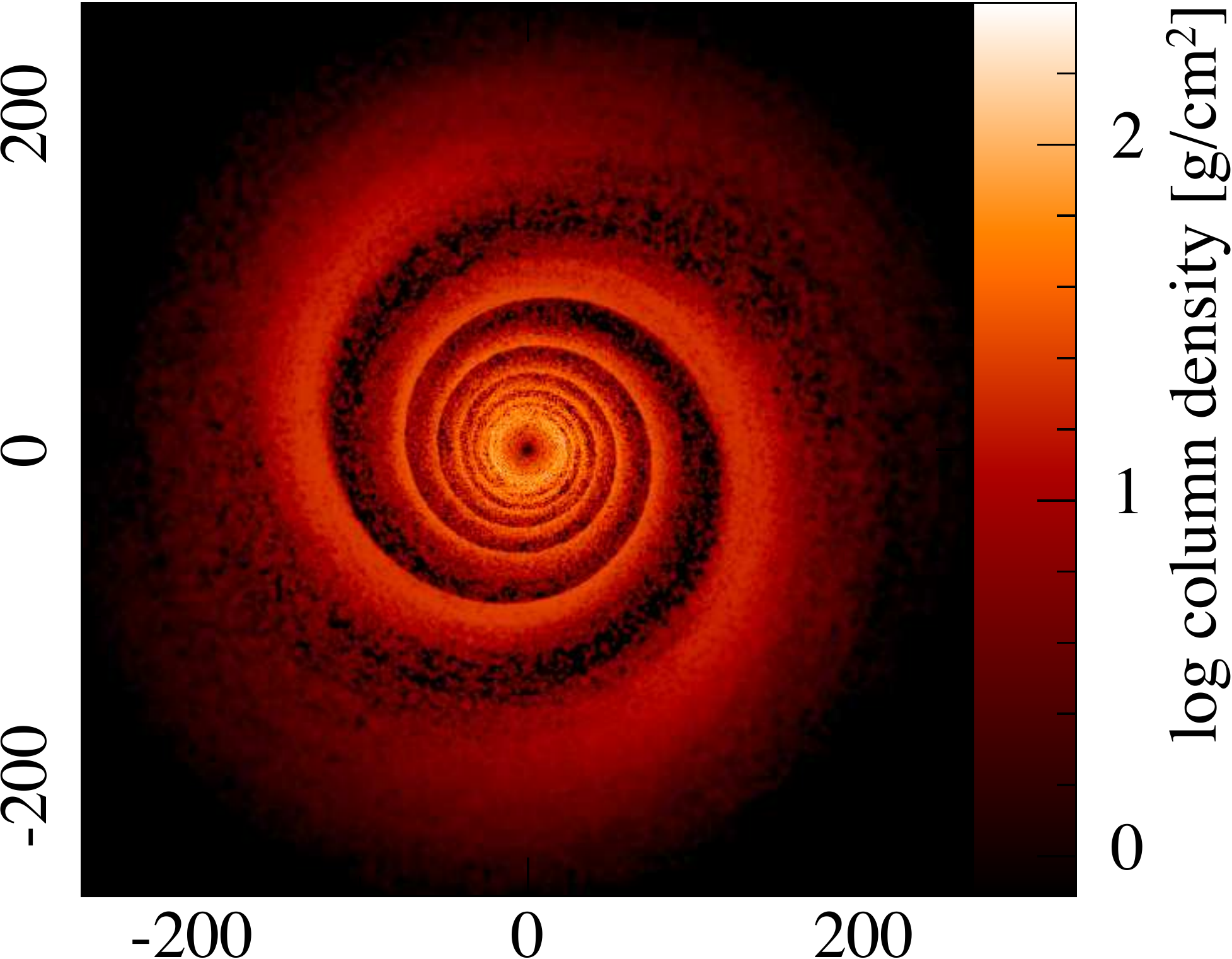} &  \includegraphics[width=0.25\textwidth,angle=90]{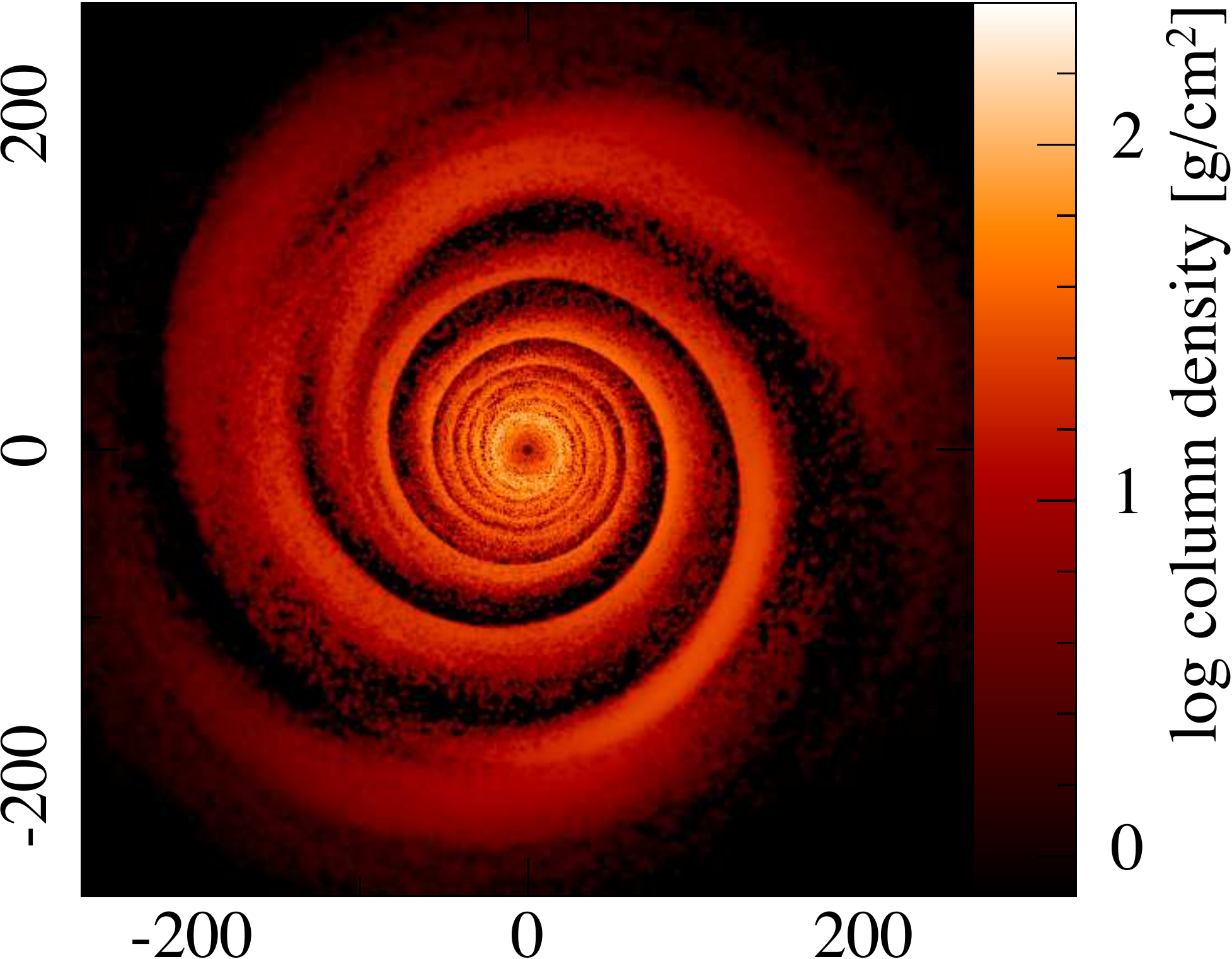}\\
    \includegraphics[width=0.33\textwidth]{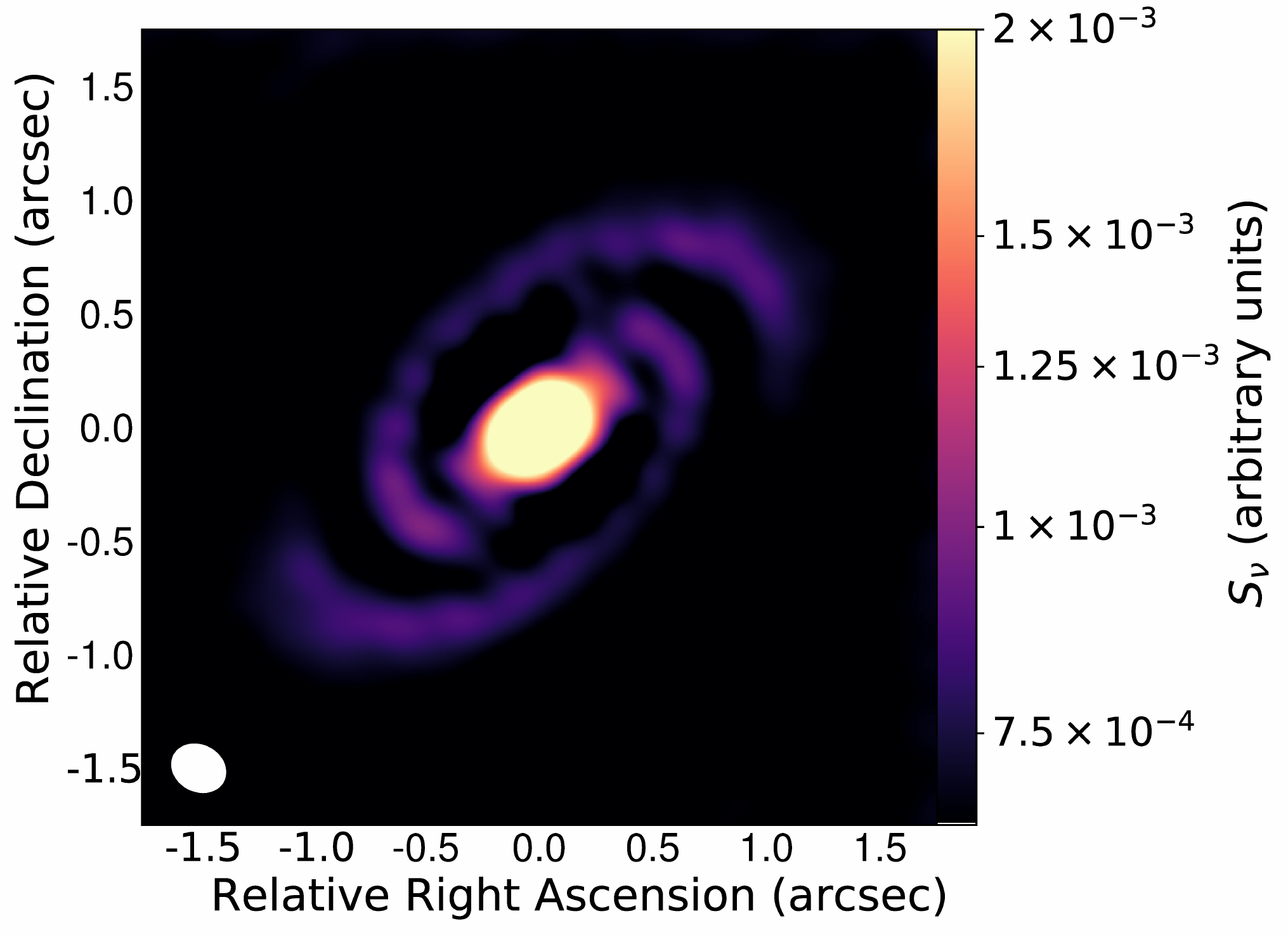} &    \includegraphics[width=0.33\textwidth]{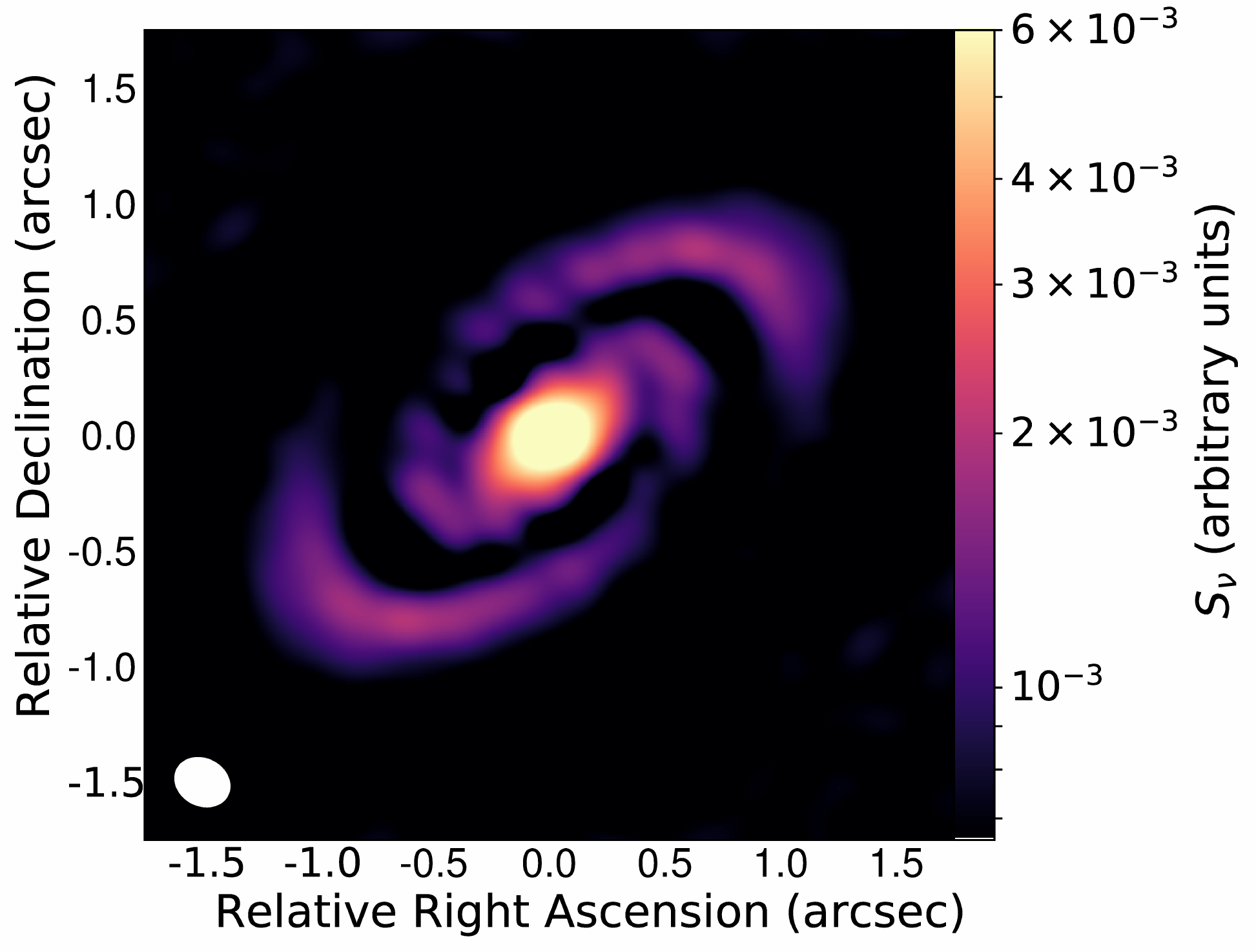} &  \includegraphics[width=0.33\textwidth]{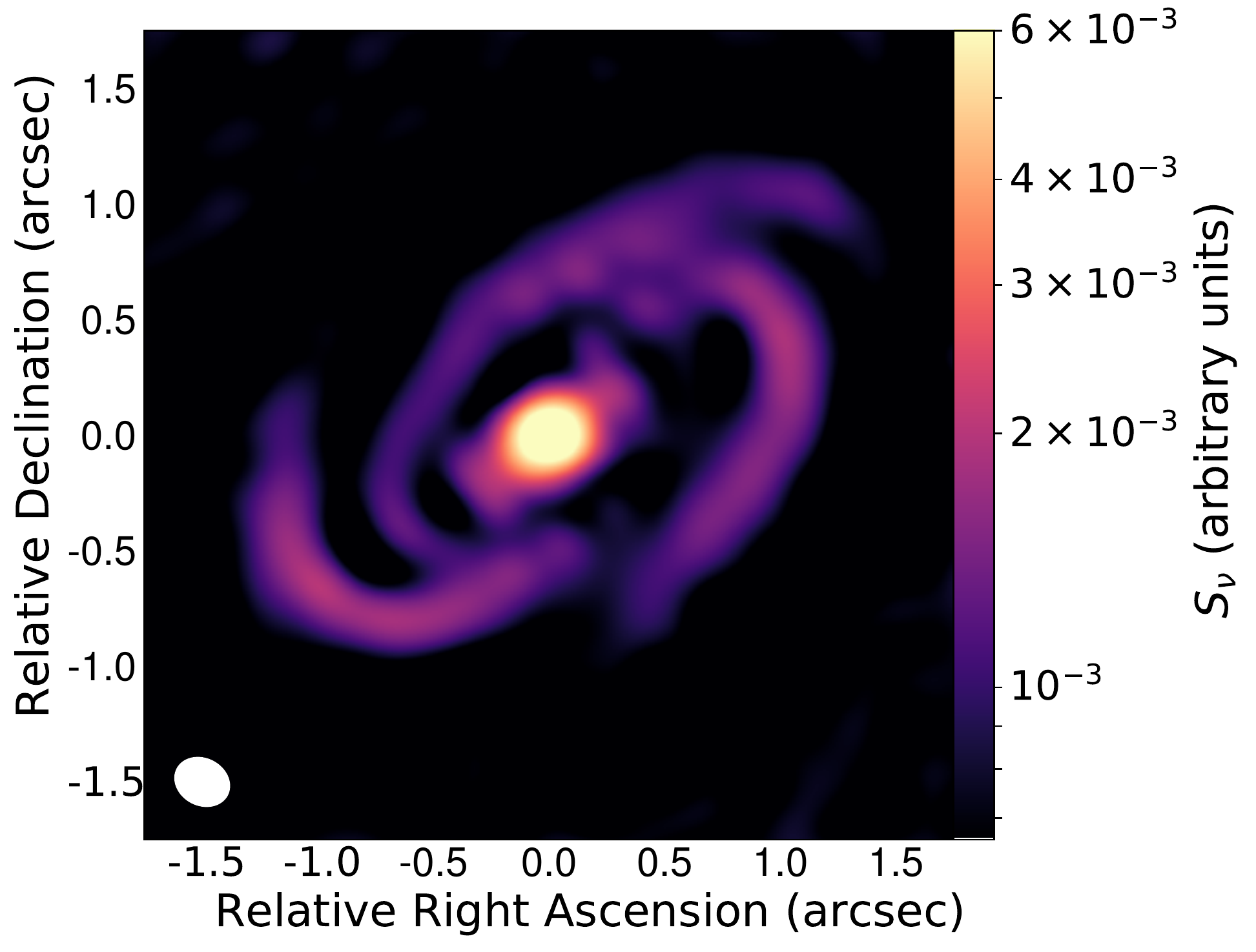}\\  
  \end{tabular}
  \caption{This Figure shows the same discs as Figure \ref{fig:highmassdiscs}, but with the spirals amplified by a factor of 1.15. In all cases, non-axisymmetric structure is now detectable.\label{fig:highmassamp}}
  \end{figure*}

We apply the simple spiral amplification technique outlined in section \ref{sec:amplify} to the discs in Figure \ref{fig:lowmassdiscs} and Figure \ref{fig:highmassdiscs}, and the results are shown in Figure \ref{fig:lowmassamp} and Figure \ref{fig:highmassamp} respectively. In Figure \ref{fig:lowmassamp}, the amplification has produced detectable spiral structure only for the discs that have the lowest minimum temperature, shown in the left and center column. For the disc irradiated at 10 K in the rightmost column, even with amplified spirals, the structure is minimal in the outer part of the disc, resulting in a disc observed with a ring-like, rather than spiral-like, morphology. This is because if the spiral amplitudes are very weak to begin with, there is very little spiral structure to actually amplify.

In the case of the larger disc masses, shown in Figure \ref{fig:highmassamp}, non-axisymmetric structure is visible in all the final disc images when we amplify the spirals. Despite the larger minimum background temperature required to prevent fragmentation, when amplified all spirals are detectable.
 This brings us to our second conclusion: accounting for the enhancement of dust at pressure maxima in the disc spiral arms considerably broadens the region of parameter space where signatures of disc self-gravity may be detected.

\section{Discussion}
In this paper, we have presented a suite of 17 SPH simulations with parameters either matching, or similar to, that of the protostellar system Elias 2-27. In all simulations, a minimum temperature floor was required to prevent the outer region of the disc fragmenting, but these temperatures are not unreasonably large. However, the requirement of a minimum temperature floor in each system to prevent fragmentation shows that there is a tenuous balance between heating and cooling, as even a small increase in the irradiative temperature may suppress the self-gravitating spirals. 

In this work, we have simulated a range of disc masses and metallicities (as a proxy for opacity) for the Elias 2-27 system, and we conclude that only in the case of the most efficient cooling (low opacity, where the spiral amplitudes are largest), is non-axisymmetric structure visible using the same observational settings as in the original detection (from \citealt{perezetal2016}), without the need to employ an enhancement of the spirals.

Since the Elias 2-27 system has an external radius of $R\sim 300$ au and is particularly massive, it will be susceptible to fragmentation without some form of external radiation \citep{riceetal2011}, highlighted by the requirement in this work of a minimum temperature floor to prevent this fragmentation. However, since the amplitude of the spirals is related to the efficiency of the cooling \citep{cossinslodatoclarke2009}, this irradiation decreases the spiral amplitude, thereby resulting in a reduced contrast between the arm and inter-arm region. Therefore, if the spirals observed in Elias 2-27 are due to gravitational instability, they only require a very small amount of support to be stable. This would suggest a relatively small level of external irradiation, since too much will suppress the spiral structure.

%


We employ a simple spiral amplification technique as a proxy to dust-trapping behaviour, and, using this, we find that spiral structure is more readily detectable in almost all cases, except where the spiral structure was initially minimal to begin with. When dust-trapping is accounted for, the region of parameter space that signatures of disc self-gravity can be detected is considerably broadened, since the intensity contrast between the arm and inter-arm region of the disc increases. This in itself is not surprising, but, what it does suggest, is that any self-gravitating protostellar disc with detectable signatures of GI in $\sim$mm emission may be indicative of dust enhancement in the spiral arms. Since the smallest grains are well coupled to the gas, this may suggest that some degree of grain growth has already occurred.

An aspect not considered in this work is infall from an envelope onto the central protostellar disc. So long as the infall rate, $\dot{M}_{\mathrm{in}}$, is  greater than the GI-driven accretion rate in the disc, $\dot{M}_{\mathrm{GI}}$, the disc will increase in mass. However, there is a maximum accretion rate due to GI that the disc can sustain, so if $\dot{M}_{\mathrm{in}} > \dot{M}_{\mathrm{max\,\, GI}}$, the disc will either fragment, or undergo violent relaxation. It has been shown that both are possible in realistic discs \citep{zhuetal2012GI}. Otherwise, if $\dot{M}_{\mathrm{in}} \leq \dot{M}_{\mathrm{max\,\, GI}}$, the disc will transport mass at the rate at which it is being fed \citep{kratteretal2008}.

Of particular interest to the results in this work is the ability of infall to drive power into lower $m$-modes in self-gravitating discs, even when the higher $m$-modes should dominate \citep{harsonoetal2011}. Furthermore, in the infall case, spirals may have amplitudes $\sim 2-5$ times higher than in an identical disc with no infall \citep{harsonoetal2011}. Therefore, the difficulty that GI alone has in replicating the observed morphology of Elias 2-27 could be somewhat mitigated by including mass infall in our calculations. However, there is no evidence for a massive envelope in the infra-red SED of Elias 2-27 \citep{andrewsetal2009}. It may, therefore, be reasonable to conclude that mass infall is not responsible for the two-armed spiral seen in the Elias 2-27 system.

Aside from the gravitational instability of the disc itself, the grand-design two armed spiral seen in Elias 2-27 may be a tidal response to a perturber, either a perturber internal to the disc in orbit around the central star, a perturber external to the disc in orbit around the central star. The latter could be in the form of a stellar encounter or flyby, which has been shown to be likely in the initial stage of disc formation \citep{bate2018}. That the spiral morphology of the Elias 2-27 system may be due to an internal or external perturber was examined by \citet{meruetal2017}, who find that an external perturber produces a reasonable match to the original observation. \citet{meruetal2017} also consider the GI origin of the spirals in the Elias 2-27 system, and find that GI can also produce a reasonable match to the original observation. 

However, GI parameters were varied over 72 simulations in the work of \citet{meruetal2017}, with one GI simulation showing a reasonable match to the original observation. The work laid out in this paper is, therefore, consistent with the results of \citet{meruetal2017}; gravitational instability can be responsible for the observed morphology of Elias 2-27, but it requires that the system falls in the narrow range of parameter space where its detection is favourable. 


It is worth noting, at this point, that the original unprocessed ALMA image may be easier to fit with a numerical model \citep{tomidaetal2017} than an image processed with the unsharped masking technique. However, it is probably reasonable to conclude that this results in a less tight constraint of disc properties. Essentially, by applying the mask, it is possible to further constrain disc properties that produced the original observation \citep{meruetal2017}.

\section{Conclusion}

We present the results of a suite of SPH simulations investigating if the morphology of the Elias 2-27 system could be due to gravitational instability. Out of a total of 17 simulations, only 6 developed GI spirals. Radiative transfer calculations were performed on each simulation to produce an intensity map, which was then synthetically observed using the \texttt{CASA} package to produce a simulated ALMA image. Finally, we applied an unsharp mask onto each image in order to enhance fainter features.

We find that only in the case of metallicity (and therefore opacity, assuming 1:1 ratio between metallicity and opacity) $0.25\times $ solar, disc-to-star mass ratio of 0.325, and a small amount of support from external irradiation (5 K) does the disc contain spiral structure evident in the final unsharped mask image. When the metallicity (opacity) of the system is set to solar value, the cooling is too inefficient to generate detectable spiral amplitudes. When the disc mass is increased, which increases the amplitude of the spirals, the system requires increased support from external irradiation in order to prevent fragmentation. As a result, the cooling rate is reduced, reducing the strength of the spirals to a level below which they are not detectable. 

However, when we amplify the spiral structure, as a proxy for dust trapping, we find that the region of parameter space in which spiral structure is detectable is broadened considerably. In this case, we detect spiral structure in the discs with $q=0.325$ and $0.25\times$ solar and $1.0$ so long as the minimum temperature satisfies $T_{\mathrm{min}}=5$ K. However, this may be problematic, since the ambient temperature in regions containing protostellar discs are significantly higher, somewhere in the region $10-30$ K \citep{hayashinakano1965,hayashi1966,larson1985,tohline1982,masunagainutsuka2000,  stamatellosetal2005a,stamatellosetal2005b,battersbyetal2014}. Although spiral structure is detectable for $T_{\mathrm{min}}=5$ K in this work, it is probably reasonable assume to such low temperatures will rarely be present in the vicinity of systems such as Elias 2-27.

Only when the disc-to-star mass ratio increases to $q\gtrsim 0.4$ is spiral structure still observable when supported with $T_{\mathrm{min}}=10$ K. The Elias 2-27 system has a measured disc-to-star mass ratio of $q\sim 0.24$ \citep{andrewsetal2009}, inferred assuming a gas-to-dust ratio of 100:1. This is not certain, and there is theoretical evidence that disc masses around young protostars have been underestimated \citep{hartmann2008}. In light of this, it is possible that the Elias 2-27 system has a $q\gtrsim 0.25$. However, this uncertainty is not unidirectional, it is also possible to overestimate disc mass, if inferring from dust measurements \citep{williamsbest2014}.

Without amplification (i.e., without dust trapping), we find that for an extended circumstellar disc such as the one in the Elias 2-27 system, the observed morphology can only be due to GI if the opacity is $\sim 0.25\times$ solar, the disc-to-star mass ratio is $q\sim0.325$ and the external irradiative temperature is $T_{\mathrm{min}}\sim 5$ K. As we have previously mentioned, it is probably unlikely that temperatures in regions neighbouring systems such as Elias 2-27 drop below 10 K. This highlights the necessity of dust-trapping in order to observe GI spirals. With amplification, the parameter space is broadened, such that the observed morphology can be explained by a disc-to-star mass ratio of  $q\gtrsim 0.4$, and an external irradiation temperature of $T_{\mathrm{min}}\lesssim 10 K$.

 In both cases, increased opacity results in decreased cooling efficiency. For the non-amplified case, this results in a decrease of the amplitude of spiral arms below the detection threshold for the original ALMA observation of the Elias 2-27 system detailed in \citet{perezetal2016}. Increasing $q$ requires an increase in $T_\mathrm{min}$ to sustain the disc against fragmentation, which may reduce the amplitude of the spirals below the detection threshold of the original ALMA observation. A decreased $q$ may push the disc out of the regime where GI is active. For GI to be responsible for the observed morphological features of the Elias 2-27 system, we conclude that $0.325 \lesssim q \lesssim 0.5$, $5 \lesssim T_{\mathrm{min}}\lesssim 10$ K, and dust-trapping should be occurring in the pressure maxima of the spiral arms.


We therefore conclude that if the spiral morphology of the Elias 2-27 system is due to gravitational instability, it must exist in a ``sweet spot'' of parameter space, where the cooling is efficient enough to generate spiral amplitudes large enough to detect, but not so large as to cause the disc to fragment, and that there is some level of external irradiation, sufficient to prevent fragmentation but not enough so that the spirals features are largely suppressed. We suggest that in light of this, other mechanisms, such as stellar encounters or flybys, should be explored to explain the observed morphology of the Elias 2-27 system.

\section*{Acknowledgements}
We thank the anonymous referee for their thoughtful review, which greatly improved the clarity of this work. We would like to thank Daniel Price for his publicly available SPH plotting code \texttt{SPLASH} \citep{splash}, which we have made use of in this paper.  CH warmly thanks Farzana Meru for elucidating discussions during the process of this work. KR gratefully acknowledges support from STFC grant ST/M001229/1. DF gratefully acknowledges support from the ECOGAL project, grant agreement 291227, funded by the European Research Council under ERC-2011-ADG. The research leading to these results also received funding from the European Union Seventh Framework Programme (FP7/2007-2013) under grant agreement number 313014 (ETAEARTH). This project has received funding from the European Research Council (ERC) under the European Union's Horizon 2020 research and innovation programme (grant agreement No 681601). This research used the ALICE2 High Performance Computing Facility at the University of Leicester. TJH acknowledges funding from Exeter's STFC Consolidated Grant (ST/M00127X/1)
%
\bibliographystyle{mnras}
\bibliography{bib}{}


\label{lastpage}

\end{document}